\documentclass[%
jmp,
 amsmath,amssymb,artice,
reprint,%
]{revtex4-1}

\usepackage{bm}
\usepackage{hyperref}
\usepackage{subcaption}
\usepackage{nicefrac}
\usepackage{graphicx}
\newcommand*{\Scale}[2][4]{\scalebox{#1}{$#2$}}

\DeclareMathAlphabet{\mathpzc}{OT1}{pzc}{m}{it}
\usepackage[utf8]{inputenc}
\usepackage[T1]{fontenc}
\usepackage{mathptmx}
\usepackage[bottom]{footmisc}

\begin{document}

\title{Fast estimation of propagation constants in crossed gratings} \thanks{This is the version of the article before peer review or editing, as submitted by an author to Journal of Optics. IOP Publishing Ltd is not responsible for any errors or omissions in this version of the manuscript or any version derived from it. The Version of Record is available online at \url{https://iopscience.iop.org/article/10.1088/2040-8986/ab6042}.}

\author{Ehsan Faghihifar}
\email{ehsan.faghihifar@alum.sharif.edu}
\affiliation{Sharif University of Technology, Azadi Ave., Tehran, 1136511155, Iran}

\author{Mahmood Akbari}
\email{makbari@sharif.edu}
\affiliation{Sharif University of Technology, Azadi Ave., Tehran, 1136511155, Iran}

\author{Seyed Amir Hossein Nekuee}
\affiliation{ICT Research Institute, North Karegar Ave., Tehran, 1439955471, Iran}

\date{\today}


\begin{abstract}
Fourier-based modal methods are among the most effective numerical tools for the accurate analysis of crossed gratings. However, leading to computationally expensive eigenvalue equations significantly restricts their applicability, particularly when large truncation orders are required. The resultant eigenvalues are the longitudinal propagation constants of the grating and play a key role in applying the boundary conditions, as well as in the convergence and stability analyses. This paper aims to propose simple techniques for the fast estimation of propagation constants in crossed gratings, predominantly with no need to solve an eigenvalue equation. In particular, we show that for regular optical gratings comprised of lossless dielectrics, nearly every propagation constant appears on the main diagonal of the modal matrix.
\end{abstract}

\maketitle

\section{Introduction}
\label{Section: Intro}

\begin{figure}[!t]
\centering
\includegraphics[width=.85\linewidth]{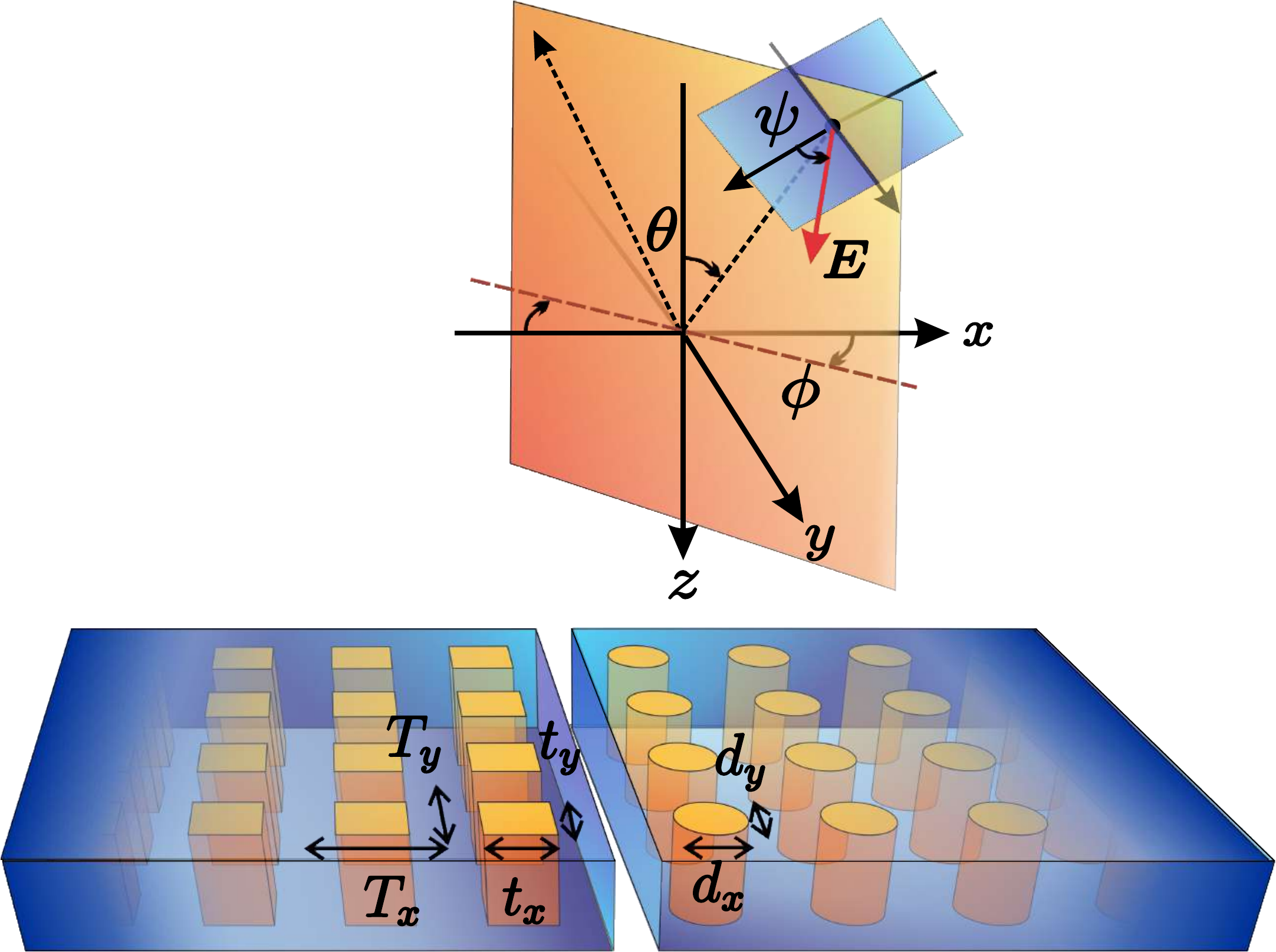}
\caption{Sample crossed gratings and their geometric parameters. The variables of an incident plane wave are also shown.}
\label{fig_3d}
\end{figure}

Periodic structures play a major role in various applications spanning the entire electromagnetic spectrum~\cite{Schurig2006, Lourtioz2008, Kildishev2013}. They either form part of the system in cases where the final structure is periodic~\cite{Ding2012, Johnson2000}, or play an intermediate role in the design of pseudo-periodic devices~\cite{Zheng2015, Aieta2012}. An important class of periodic structures fall under the category of crossed gratings. In general, crossed gratings are two-dimensional periodic arrays wherein the composition is kept constant over the thickness~\cite{Li1997}. Schematics of two simple crossed gratings are shown in Fig.~\ref{fig_3d}. The significance of these structures arises in part due to their compatibility with conventional planar fabrication techniques. As a result, a large portion of the periodic and pseudo-periodic devices for various applications such as absorbers \cite{Landy2008}, waveguides \cite{Christ2003}, phase shifters \cite{Antoniades2003}, polarization converters \cite{Grady2013}, and antennas \cite{LeWei2010} are in general crossed grating structures. Therefore, fast, reliable, and accurate methods for characterizing their electromagnetic properties are of great interest.

While fully numerical methods like the finite elements method~\cite{Bao2005} and the finite-difference time domain method~\cite{Correia2004} can be used in the analysis of these structures, they usually need mesh sizes that are significantly smaller than the wavelength, in order to obtain acceptable levels of accuracy. This would in turn increase their execution time and limit their applicability significantly. Other numerical methods such as the integral equation method~\cite{Dobson1991}, the boundary variation method~\cite{Bruno1993}, and the rigorous coupled wave analysis (RCWA) technique~\cite{Moharam1995} are also used in simulation of crossed gratings. Nevertheless, all of them are computationally expensive, especially for materials with high refractive indices, high index contrasts, or complex structures. On the other hand, RCWA still represents one of the most significant categories of semi-analytical methods for the analysis of periodic media. Although a modified reformulation of RCWA has been proposed as the Fourier modal method (FMM)~\cite{Li1997} to effectively improve the convergence rate, most of the run time is still spent on solving an inherent eigenvalue equation. The eigenvalues, i.e. the longitudinal propagation constants of the grating, play a key role in applying the boundary conditions at the interface of layers~\cite{Li1996-2}. Moreover, they can be important in stability or convergence analysis of the numerical methods. Since the resultant equation could be immense in practice, the plausible existence of an asymptotic approximation could be of great value in modal analyses of grating structures.

This paper presents simple estimation schemes for characterization of the propagation constants in a crossed grating. We have uncovered this fascinating property that propagation constants of a grating make a pattern, and that pattern can be expressed as a weighted summation over the corresponding patterns of uniform layers with the same dielectric constants. Moreover, a relationship between the diagonal entries and the eigenvalues of the modal matrix is specified, which can be used along with the sequential pattern completion property, to provide a universal estimation scheme, covering a variety of grating problems. In Sec.~\ref{Section: Math} of this paper, the governing equation of the propagation constants is extracted. In Sec.~\ref{Section: Asymp}, basic definitions are presented and estimation schemes are developed. In Sec.~\ref{Section: Num}, the propounded ideas are clarified and numerically illustrated. Finally in Sec.~\ref{Section: Conc}, a summary of the work is provided.

\section{Mathematical Foundation}
\label{Section: Math}

Here, we present a formulation of the grating problem and extract the governing equation of the propagation constants through direct manipulation of the wave equation. Throughout the text, matrices and vectors are denoted by bold letters. Consider a two-dimensional periodic dielectric structure, which is lossless and non-magnetic. The structure is positioned in the $xy$-plane, uniformly stretched along the $z$-axis, with the relative permittivity matrix $\Scale[0.9]{\epsilon(x,y)}$. Let $\Scale[.95]{\Lambda_x}$ and $\Scale[.95]{\Lambda_y}$ be periods along $x$ and $y$-axes, so that $\Scale[0.85]{\epsilon (x+\Lambda_x,y) =\epsilon( x,y +\Lambda _y) =  \epsilon(x,y)}$. If we define $\Scale[0.9]{\mathcal{J}_n = \{-n,...,0,...,n\} }$ and $\Scale[0.9]{\mathcal{K}_n=\{1,2,...,n\} }$ for any positive integer $n$, the dielectric function could be expanded as below:

\begin{eqnarray}\label{Eq-1}
\Scale[.9]{
\begin{array}{l}
 \epsilon(x,y) = {\displaystyle \sum\limits_{m,n} { {{{ \epsilon }^{m,n}}{e^{ - jmx({{2\pi }}/{\Lambda _x})}}{e^{ - jny({2\pi}/{\Lambda_y})}}}}},
\\
{{{ \epsilon }^{m,n}}} = \frac{1}{{\Lambda _x}{\Lambda _y}}{\displaystyle \iint\limits_{\Scale[.65]{\rm{unit\hspace{2pt} cell}}}{\epsilon(x,y){e^{ jmx({{2\pi }}/{{{\Lambda _x}}})}} {e^{ jny({{2\pi }}/{{{\Lambda _y}}})}}}\,dxdy}.
\end{array}
}
\end{eqnarray}

In which $\Scale[0.9]{m \in \mathcal{J}_M}$ and $\Scale[0.9]{n \in \mathcal{J}_N}$, for the truncation orders $\Scale[.95]{M}$ and $\Scale[.95]{N}$. The total number of Fourier orders or simply the total order, will consequently be $\Scale[0.9]{S=(2M+1)(2N+1)}$. Since the medium is periodic along $x$ and $y$ and uniform in the $z$ direction, a Floquet representation for the electric field can be utilized. If $\Scale[.95]{k_{x0}}$ and $\Scale[.95]{k_{y0}}$ are transverse wave numbers of the incident plane wave, every discrete Floquet wave vector has a form $\Scale[0.9]{ {\bm k_{m,n}}=(k_{xm},k_{yn},k_z)}$, in which $\Scale[0.9]{ k_{xm}=k_{x0}+2m\pi / \Lambda_x }$, $\Scale[0.9]{ k_{yn}=k_{y0}+2m\pi / \Lambda_y }$, and the longitudinal propagation constant $\Scale[.95]{k_z}$ is to be found accordingly. The electric field could be expanded in the subsequent form:

\begin{eqnarray}\label{Eq-2}
\Scale[.9]{
\begin{array}{l}
\bm E(\bm r) = {\displaystyle \sum\limits_{m,n} {{{\bm E}_{m,n}}(z){e^{-j{k_{yn}}y}}} {e^{-j{k_{xm}}x}}} =\\\hspace{28pt}
\frac{1}{2\pi}{\displaystyle \int\nolimits_{ - \infty }^\infty { \sum\limits_{m,n} {{{e^{-j{k_{yn}}y}}} } {e^{-j{k_{xm}}x}}{{{\bm E}_{m,n}}}(k_z){e^{ - j{k_z}z}}dk_z}}.
\end{array}
}
\end{eqnarray}

Now, if we write the wave equation for the electric field as below, in which $\Scale[.95]{k_0}$ is the free space wave number:

\begin{equation}\label{Eq-3}
\Scale[.9]{
\bm\nabla  \times \bm\nabla  \times \bm E(\bm r) - k_0^2 \epsilon(x,y)\bm E(\bm r) = \bm{0}.
}
\end{equation}

By substituting Eq.~\ref{Eq-1} and Eq.~\ref{Eq-2} in Eq.~\ref{Eq-3} and eliminating the integrals, the wave equation for the electric field could be expanded as following:

\begin{eqnarray}\label{Eq-4}
\Scale[.9]{
\begin{array}{l}
{\displaystyle \sum\limits_{m,n} { {{e^{ - j{k_{yn}}y}}} } {e^{ - j{k_{xm}}x}}{e^{ - j{k_z}z}}{{{\bm k}_{m,n}}} \times ({{{\bm k}_{m,n}}} \times{{{\bm E}_{m,n}}}( {{k_z}} ))} + \\
k_0^2{\displaystyle\sum\limits_{{ {m',n'}\atop {m,n} \hfil\hfil}} {{ {{{{ \epsilon }^{m',n'}}} }{e^{ - j{k_{x( {m + m'})}}x}}{e^{ - j{k_{y( {n + n'} )}}y}}{e^{ - j{k_z}z}}} } {{{\bm E}_{m,n}}}( {{k_z}})}  = \bm{0}.
\end{array}
}
\end{eqnarray}

Since the exponential terms are linearly independent, coefficients of the corresponding exponential terms can be all set to zero. Thus, a set of $\Scale[.95]{S}$ vector equations is obtained for all index pairs $\Scale[.95]{(m,n)}$, which gives $\Scale[.95]{3S}$ scalar equations. Henceforth, we would rather replace normalized wave vectors and wave numbers $\Scale[.9]{{{\bm N_{m,n}}} = {{\bm k_{m,n}}}/{k_0}}$, $\Scale[0.9]{ N_{xm}=k_{xm}/k_0$, $\Scale[.9]{N_{yn}=k_{yn}/k_0}}$, and $\Scale[0.9]{ \lambda=k_z/k_0}$. The values of $\lambda$ are the normalized longitudinal propagation constants of the grating and the eigenvalues of the problem. Now, Eq.~\ref{Eq-4} can be expressed as a set of equations for each index pair $\Scale[.95]{({m,n})}$, as following:

\begin{equation}\label{Eq-5}
\Scale[.9]{
{{{\bm N}_{m,n}}} \times ({{{\bm N}_{m,n}}} \times {{{\bm E}_{m,n}}}( \lambda )) + {\displaystyle \sum\limits_{m',n'}  {{{ \epsilon }^{m-m',n-n'}}} {{{\bm E}_{m',n'}}}( \lambda )} = \bm{0}.
}
\end{equation}

In a 2D grating problem, in order to specify the dominant Fourier products and arrange the equations in the matrix form, one needs to define a bijective relation $\Scale[0.85]{\nu:{\mathcal{J}_M} \times {\mathcal{J}_N} \longleftrightarrow  {\mathcal{K}_S}}$ so that every index pair $\Scale[.95]{(m,n)}$ corresponds to some unique ordinal number $\Scale[0.95]{ s=\nu ( {m,n} )}$ as the eigenvalue index. Hereafter, we will call such a relationship an order function. Now we rearrange the set of equations specified in Eq.~\ref{Eq-5} into a block matrix equation, by separating the coefficients of each $x, y$, or $z$ components. Initially, a set of matrices and vectors are defined elementwise as following, for all $\Scale[0.85]{ i,j \in {\mathcal{K}_S}\hspace{2pt}}$ and $\Scale[0.9]{ u \in \{x,y,z\}}$:

\begin{eqnarray}\label{Eq-6}
\Scale[.9]{
\begin{array}{l}
[\bm{F}]_{i,j} =  {\epsilon ^{{m,n}}},\hspace{55pt}( {m,n}) = {{\nu}^{ - 1}}( i) - {{\nu}^{ - 1}}( j),\\
\left[\bm N_x\right]_{i,j} = {\delta _{ij}}{N_{xm}},\hspace{40pt}( {m,n}) = {{\nu}^{ - 1}}( i ),\\
\left[{{\bm N}_y}\right]_{i,j} = {\delta _{ij}}{N_{yn}},\hspace{42pt}( {m,n}) = {{\nu}^{-1}}(i),\\
\left[{{\bm e}_u}\right]_i = {{{\bm E}_{m,n}^{u}}},\hspace{56pt}( {m,n}) = {{\nu}^{ - 1}}( i ).
\end{array}
}
\end{eqnarray}

In which $\delta _{ij}$ is the Kronecker delta. For further manipulations, we will also define the following block matrices, in which $\Scale[.85]{\bm{Q} = \bm{F}- {\bm N_x^2} - {\bm N_y^2}}$ is a square matrix and will have a decisive role in our analysis:

\begin{equation}\label{Eq-7}
\Scale[.82]{
\begin{array}{l}
{\bm N} = \left[ \begin{array}{l}
{{{\bm N}_x}}\\
{{{\bm N}_y}}
\end{array} \right],\hspace{2pt} \bm{{ e}_t} = \left[ {\begin{array}{*{20}{c}}
{\bm{{ e}_x}}\\
{\bm{{ e}_y}}
\end{array}} \right],\hspace{2pt}\bm{F}_2 = \left[ {\begin{array}{*{20}{c}}
{\bm F}&{\bm 0}\\
{\bm 0}&{\bm F}
\end{array}} \right]
\end{array},\hspace{2pt}\bm{Q}_2 = \left[ {\begin{array}{*{20}{c}}
{\bm Q}&{\bm 0}\\
{\bm 0}&{\bm Q}
\end{array}} \right].
}
\end{equation}

Now, the set of equations in Eq.~\ref{Eq-5} may be restated in the matrix form as below, wherein the prime symbol $'$ denotes the conjugate transpose of a matrix. Note that $\Scale[.9]{\bm N_x}$ and $\Scale[.9]{\bm N_y}$ are real diagonal matrices and $\Scale[.9]{\bm F = \bm F'}$, for lossless gratings.

\begin{equation}\label{Eq-8}
\Scale[.85]{
\left[ {\begin{array}{*{20}{c}}
{\bm{Q_2} + \bm{N}\bm{N}'- {\lambda ^2} {\bm{I}}}&{\lambda {\bm{{ N}}}}\\
{\lambda {\bm{N}'}}&{\bm{Q}}
\end{array}} \right]\left[ {\begin{array}{*{20}{c}}
{\bm{{ e}_t}}\\
{\bm{{ e}_z}}
\end{array}} \right] = {\bm{0}}.
}
\end{equation}

Eq.~\ref{Eq-8} is a quadratic eigenvalue equation in terms of $\lambda$, and should be treated accordingly \cite{gohberg1982matrix}. In polynomial eigenvalue problems, the singularity of coefficients may considerably affect the properties of the solutions \cite{Tisseur2001}. It is straightforward to see in Eq.~\ref{Eq-8}, $\Scale[.95]{2S}$ of the total $\Scale[.95]{6S}$ roots approach infinity, and the rest are the desired normalized propagation constants. In order to remove the nullity, ${\bm {e_z}}$ should be omitted from Eq.~\ref{Eq-8}. If we expand Eq.~\ref{Eq-8}, omit ${\bm {e_z}}$, and finally use Woodbury matrix identity \cite{golub2012matrix} to simplify the result, the governing eigenvalue equation is obtained as following:

\begin{eqnarray}\label{Eq-9}
\Scale[.9]{
\begin{array}{l}
\left( { {\bm{P}} - {\lambda^2} {\bm{I}}} \right)\bm{{e}_t} = \bm{0},\\
\bm{P} = \bm{Q}_2 + \bm{N}(\bm{N}' - {\bm F}\bm{N}'{\bm{F}_2^{-1}}).
\end{array}
}
\end{eqnarray}

As it turned out, eigenvalues of the matrix $\Scale[.95]{\bm{P}}$ are the squared propagation constants of the grating. The matrix $\Scale[.95]{\bm{P}}$, which will be called the modal matrix, is expressed as the sum of two terms. We will see later how the two summands contribute in characterizing the eigenvalues. Eventually, let us point out a characteristic property of the propagation constants in a lossless isotropic grating. It can be shown that the propagation constants are either pure real or pure imaginary numbers. In other words, solutions of Eq.~\ref{Eq-9} satisfy the following:

\begin{equation}\label{Eq-10}
\Scale[.9]{
{\rm Im}({\lambda^2})=0.
}
\end{equation}

Notice that the modal matrix $\Scale[.95]{\bm P}$ in Eq.~\ref{Eq-9} is not Hermitian, hence the above statement is not trivial. To prove it, consider the first element in the matrix product of Eq.~\ref{Eq-8}, and multiply it from the left by $\Scale[.95]{\bm{e_t}'}$ to obtain the following equation:

\begin{equation}\label{Eq-11}
\Scale[.9]{
\begin{array}{l}
\bm{e_t}'(\bm{Q_2} + \bm{N}\bm{N}'- \lambda^2 \bm{I})\bm{e_t} + \lambda {\bm{e_t}'}\bm{N}{\bm{e_z}} = 0.
\end{array}
}
\end{equation}

On the other hand, from the second element in Eq.~\ref{Eq-8}, one can easily obtain $\Scale[.85]{{\bm{e_t}'}\bm{N} = -{\bm{e_z}'}\bm{Q}/\lambda'}$ where $\Scale[.85]{\lambda \ne 0}$, otherwise Eq.~\ref{Eq-10} obviously holds. Also note that $\Scale[.95]{\bm Q}$ and $\Scale[.95]{\bm {Q_2}}$ are Hermitian matrices. Now, by replacing this expression in Eq.~\ref{Eq-11}, the following result can be obtained:

\begin{equation}\label{Eq-12}
\Scale[.9]{
\begin{array}{l}
\bm{e_t}'\bm{Q_2}\bm{e_t} + \|\bm{e_t}'\bm{N}\|^2 - \lambda^2(\|\bm{e_t}\|^2 + {\bm{e_z}' \bm{Q}\bm{e_z}}/{\left|\lambda\right|^{2}}) = 0.
\end{array}
}
\end{equation}

The above equation has the form of $\Scale[.85]{a-\lambda^2 b=0}$, with real coefficients $\Scale[.95]{a}$ and $\Scale[.95]{b}$. It follows immediately that $\Scale[.9]{\lambda^2}$ is a real number, satisfying Eq.~\ref{Eq-10}. We have excluded the exceptional case where $\Scale[.9]{a=b=0}$ and made use of the fact that for any Hermitian matrix $\Scale[.95]{\bm M}$ and vector $\Scale[.9]{\bm v}$, the expression $\Scale[.9]{\bm{v}'\bm{M}\bm{v}}$ is real-valued.

\section{Estimation Scheme}
\label{Section: Asymp}

\subsection{Basic Definitions}
\label{Subsec: Def}

Before proceeding with our analysis, it will be beneficial for the sake of clarity and brevity to give some definitions. Initially, let us recall the definition of a "multi-set", being a generalized set in which the repetition of elements is allowed. For a square matrix $\Scale[0.95]{\bm{A}}$ and a positive integer $n$, the roots of $\Scale[0.9]{ {\rm{ det}}( {\bm A} - x^n{\bm I})}$ and the diagonal entries of $\Scale[.95]{\bm A}$, both characterize typical multi-sets, which we denote by $\Scale[0.9]{ \bm{\sigma}_n(\bm A)}$ and $\Scale[0.9]{ {\bm\delta}(\bm{A})}$, respectively. Note that $\Scale[0.9]{ {\bm{\sigma}_1 }(\bm{A})={\bm \sigma}(\bm{A})}$ would simply denote the eigenvalue spectrum of the matrix $\Scale[0.95]{\bm A}$. The $\bm\delta$-notation shouldn't be mixed up with the Kronecker delta, which always precedes a subscript. All multi-sets in this paper are composed of complex numbers, hence it is possible to sort them in descending order with respect to real parts of their elements. Elements with equal real parts are arbitrarily sorted. For a multi-set $\Scale[0.85]{ \bm{\mathcal{M}}}$ and a positive integer $n$, let $\Scale[.8]{\left\langle \bm{\mathcal{M}}\right\rangle_{n}}$, $\left\{\Scale[.8]{\bm{\mathcal{M}}}\right\}_{\Scale[.65]{n}}^{\Scale[.55]{\geqslant}}$, and $\left\{\Scale[.8]{\bm{\mathcal{M}}}\right\}_{\Scale[.65]{n}}^{\Scale[.55]{\leqslant}}$ denote respectively to the $n$-th element of $\Scale[0.85]{ \bm{\mathcal{M}}}$, the multi-set composed of the first $n$ elements of $\Scale[0.85]{ \bm{\mathcal{M}}}$, and the multi-set composed of the last $n$ elements of $\Scale[0.85]{ \bm{\mathcal{M}}}$. We would also like to define the powers of a multi-set in two different ways. Let $\Scale[0.8]{ {\bm{ \mathcal{M}}^{n}}}$ denote a multi-set in which every element of $\Scale[0.85]{ \bm {\mathcal{M}}}$ is replicated $n$ times, and $\Scale[0.8]{ {\bm{ \mathcal{M}}^{[n]}}}$ denote a multi-set in which every element of $\Scale[0.85]{ \bm{\mathcal{M}}}$ is powered by $n$.

In terms of multi-set notations, solutions of Eq.~\ref{Eq-9} or the propagation constants satisfy $\Scale[0.9]{ \lambda \in {\bm\sigma_2 }(\bm{P})}$. Since we are more interested in squared propagation constants, i.e. $\Scale[.95]{\lambda^2}$, let us define $\Scale[0.85]{ \bm{\mathcal{C}}_{\epsilon}={\bm{\sigma}_2 }(\bm{P} )^{[2]}}$ to be comprised of all $\Scale[.95]{4S}$ squared propagation constants. It is easy to check the simpler relationship $\Scale[0.85]{ \bm{ \mathcal{C}}_{\epsilon}={\bm\sigma}({\bm P})^2}$ holds. Now, we will sort the propagation constants and define some functions accordingly. Let the $s$-th propagation constant be denoted by $\Scale[.95]{\lambda_s}$, such that $\Scale[.85]{\lambda_s^2 = \left\langle\bm{\mathcal{C}}_{\epsilon}\right\rangle_{s}}$ for $\Scale[0.85]{ s \in {\mathcal{K}}_{4S}}$. Now, we define $\Scale[0.85]{ {\rm g}(s)={\rm Re}(\lambda_s ^2)}$ and $\Scale[0.85]{ {\rm{g^*}}(s) = {\rm Im}({\lambda_s ^2})}$. In the previous section we showed that for a lossless isotropic grating $\Scale[.85]{{\rm{g^*}}(s) = 0}$ and $\Scale[.85]{{\rm{g}}(s)=\lambda_s^2}$. We will call the plot of $\Scale[0.9]{ {\rm g}(s)}$ against the ordinal index $s$ the "eigenvalue pattern". Furthermore, it will be shown that for large enough truncation orders, $\Scale[0.9]{{\rm{g}}( s )}$ pursues a particular piecewise-linear pattern, which we call the "asymptotic pattern" and denote by $\Scale[0.9]{{\rm{\tilde g}}( s )}$. The word "asymptotic", here and everywhere in our context refers to a limiting approximation with a bounded pointwise error, i.e. $\Scale[0.85]{ \mathop {\lim }\nolimits_{S \to \infty }{\left\| {\rm{g} - \rm{\tilde g}} \right\|_\infty } < \infty }$, in which $\Scale[0.85]{\left\| \cdot \right\|_\infty }$ is the maximum norm. It will be shown later that for a grating problem, the asymptotic pattern can be characterized by the structural parameters only, needless of solving the eigenvalue equation.

\subsection{Uniform Layers}
\label{Subsec: Unif}

Before moving on with our analysis for a general grating, let us study the special case of a uniform layer with the dielectric constant $\Scale[.95]{\epsilon _1}$. Here, the Fourier coefficient matrix will be $\Scale[0.9]{ \bm{F} = {\epsilon _1}\bm{I}}$, resulting in the second summand in Eq.~\ref{Eq-9} to vanish and $\Scale[.9]{\bm P = \bm{Q_2}}$. In terms of eigenvalue spectra, it means that $\Scale[0.85]{ \bm{ \mathcal{C}}_{\epsilon}=\bm\sigma( \bm Q)^{4}}$. In order to obtain the eigenvalue pattern, the elements of $\Scale[.85]{\bm{ \mathcal{C}}_{\epsilon}}$ should be sorted and plotted versus the ordinal index $s$. Since $\Scale[.9]{\bm Q}$ is a diagonal matrix, the eigenvalues can easily be specified as following, in which $\Scale[.95]{\lambda_s^2}$ is the $s$-th eigenvalue in order:

\begin{equation}\label{Eq-13}
\Scale[0.9]{
\begin{array}{l}
\bm\sigma ( \bm Q ) = \bm\delta ( \bm Q )=  \bm\delta ( {\epsilon_1}\bm{I} - {{\bm N_x^2} - {\bm N_y^2}} ) = \\
\hspace{30pt} {\left\{\lambda_s^2 := {{\epsilon_1} - {{( {{N_{x0}} + m/{T_x}} )}^2} - {{( {{N_{y0}} + n/{T_y}} )}^2}} \right\}_{m,n}}.
\end{array}
}
\end{equation}

Each element of $\Scale[0.85]{ \bm\sigma( \bm Q)}$ with $\Scale[0.9]{ s \in {\mathcal{K}}_S}$, defines a characteristic ellipse in the $mn$-plane. These ellipses are concentric at $\Scale[0.75]{( {-N_{x0}{T_x}},{-N_{y0}{T_y}})}$, with the semi-major and semi-minor axes $\Scale[0.85]{ r_x={T_x}({ {\epsilon _1} - \lambda _{s}^2})^{\nicefrac{1}{2}}}$ and $\Scale[0.85]{ r_y={T_y}({ {\epsilon _1} - \lambda _{s}^2})^{\nicefrac{1}{2}} }$. Counterintuitively, eigenvalues are dependent on the characteristics of the incident wave, however, since the surrounding environment is vacuum, the center of the ellipses is located close to the origin and the relationship with the incident wave is practically insignificant. As a result, the eigenvalue pattern of a uniform layer, i.e. $\Scale[0.85]{ {\rm {g_1}}( s )={\rm {g}}( s )|_{{\epsilon (x,y)}={\epsilon_1}}}$ can be estimated as following:

\begin{equation}\label{Eq-14}
\Scale[0.85]{
{\rm g}_{1}(s) = \lambda _{s}^2 \approx {\epsilon_1} - (m/{T_x})^2 - (n/{T_y} )^2, \hspace{30pt}({m,n}) = {{\gamma}^{-1}}(s)}.
\end{equation}

In which $\Scale[0.85]{\gamma:{\mathcal{J}_M} \times {\mathcal{J}_N} \longleftrightarrow {\mathcal{K}_S}}$ is an order function to sort the eigenvalues of Eq.~\ref{Eq-13}; described as following:

\begin{equation}\label{Eq-14.5}
\Scale[.8]{
\begin{array}{l}
\gamma({m,n})= {\#} \{ (p,q) \in {\mathcal{J}_M} \times {\mathcal{J}_N} \mid \\
\hspace{15pt} ({{N_{x0}} + {p}/{T_x}})^2  + ({{N_{y0}} + {q}/{T_y}})^2  \leqslant
({{N_{x0}} + {m}/{T_x}})^2 + ({{N_{y0}} + {n}/{T_y}})^2\}.
\end{array}
}
\end{equation}

Note that the center point $\Scale[0.75]{( {-N_{x0}{T_x}},{-N_{y0}{T_y}})}$ can always be modified insignificantly to assure no characteristic ellipse passes through more than one lattice point, and the above relationship remains bijective. Now, in order to provide a visual insight into the asymptotic behavior of the propagation constants, we'd like to omit $\Scale[.9]{\gamma(m,n)}$ from the above equation and estimate $\Scale[0.9]{\lambda _{s}^2}$ in terms of the ordinal index $s$ only. Since the $s$-th element of $\Scale[0.9]{ \bm{\mathcal{C}}_{\epsilon}}$ corresponds to the $\nicefrac{s}{4}$-th eigenvalue of $\Scale[0.9]{ \bm\sigma(\bm Q )}$, the area of the corresponding ellipse approximately equals to the number of enclosed lattice points, i.e. $\nicefrac{s}{4}$ by definition. On the other hand, the area can be calculated by $\Scale[0.9]{\pi{r_x}{r_y}}$. Hence the following asymptotic relationship holds for $\Scale[.9]{S \to \infty}$:

\begin{equation}\label{Eq-15}
\Scale[.95]{
s/4 \approx \pi {r_x}{r_y}{\hspace{10pt}}\Rightarrow{\hspace{10pt}}\lambda _s^2 \approx  {\epsilon _1} - {( {4\pi {T_x}{T_y}})^{ - 1}}s.
}
\end{equation}

\begin{figure}[!t]	
	\centering
	\begin{subfigure}{0.5\textwidth}
		\centering
		\includegraphics[width=0.8\linewidth]{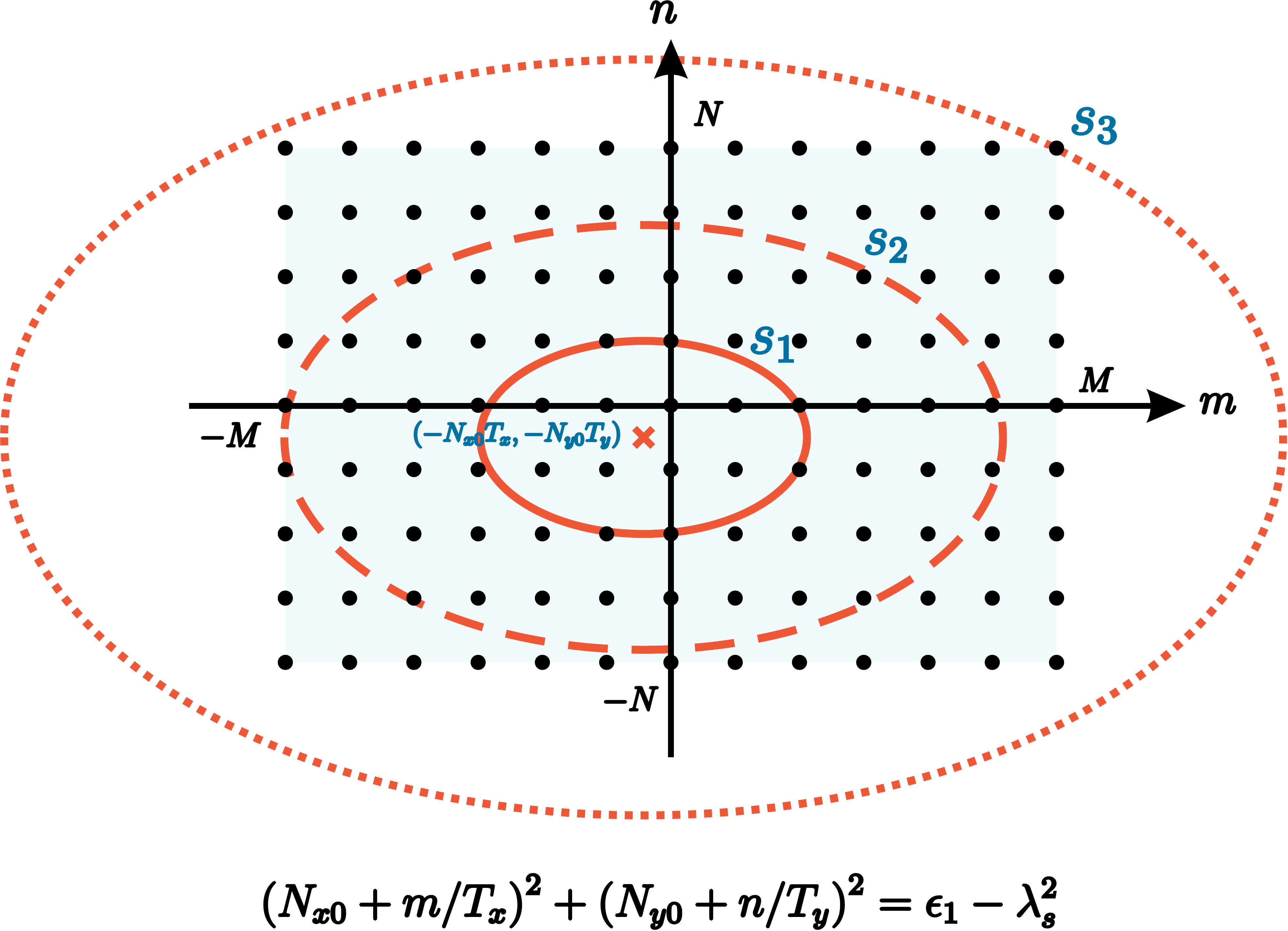}
		\caption{}\label{fig_ellip_a}		
	\end{subfigure}
\quad
	\begin{subfigure}{0.5\textwidth}
		\centering
		\includegraphics[width=0.8\linewidth]{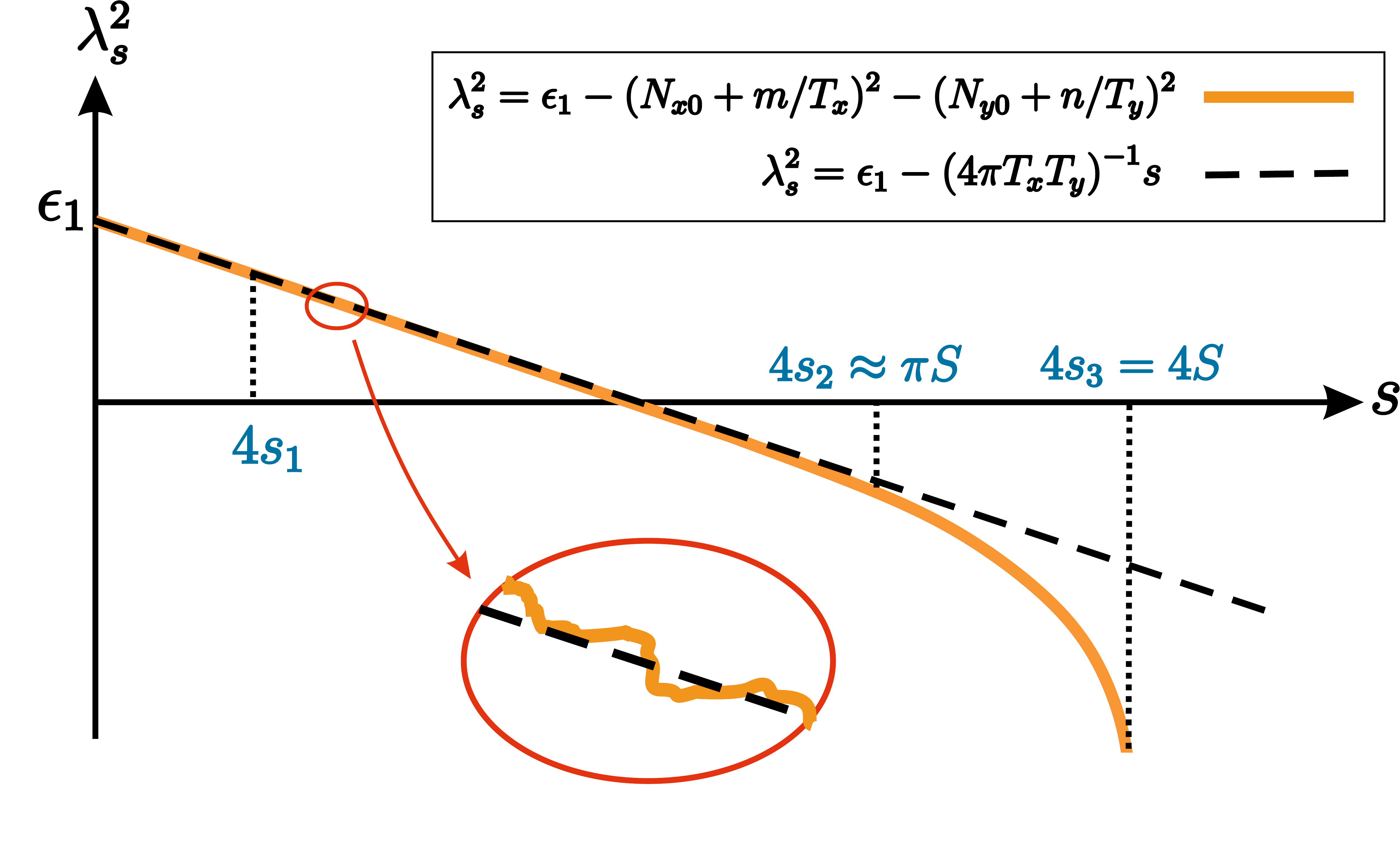}
		\caption{}\label{fig_ellip_b}
	\end{subfigure}

	\caption{(\subref{fig_ellip_a}) Characteristic equation and the corresponding ellipses for a uniform layer of $\Scale[.95]{\epsilon_1}$, located on the integeral lattice of Fourier orders; ellipses with solid, dashed, and dotted lines, correspond to eigenvalues within the linear region ($s_1$), at the edge of the linear region ($s_2$), and last eigenvalue ($s_3$), respectively. (\subref{fig_ellip_b}) The eigenvalue pattern and the asymptotic pattern of the uniform layer $\Scale[.95]{\epsilon_1}$; the number of propagation constants is four times the number of lattice points.}
	\label{fig_ellip}
\end{figure}

\begin{figure}[!t]
\centering
\includegraphics[width=.95\linewidth]{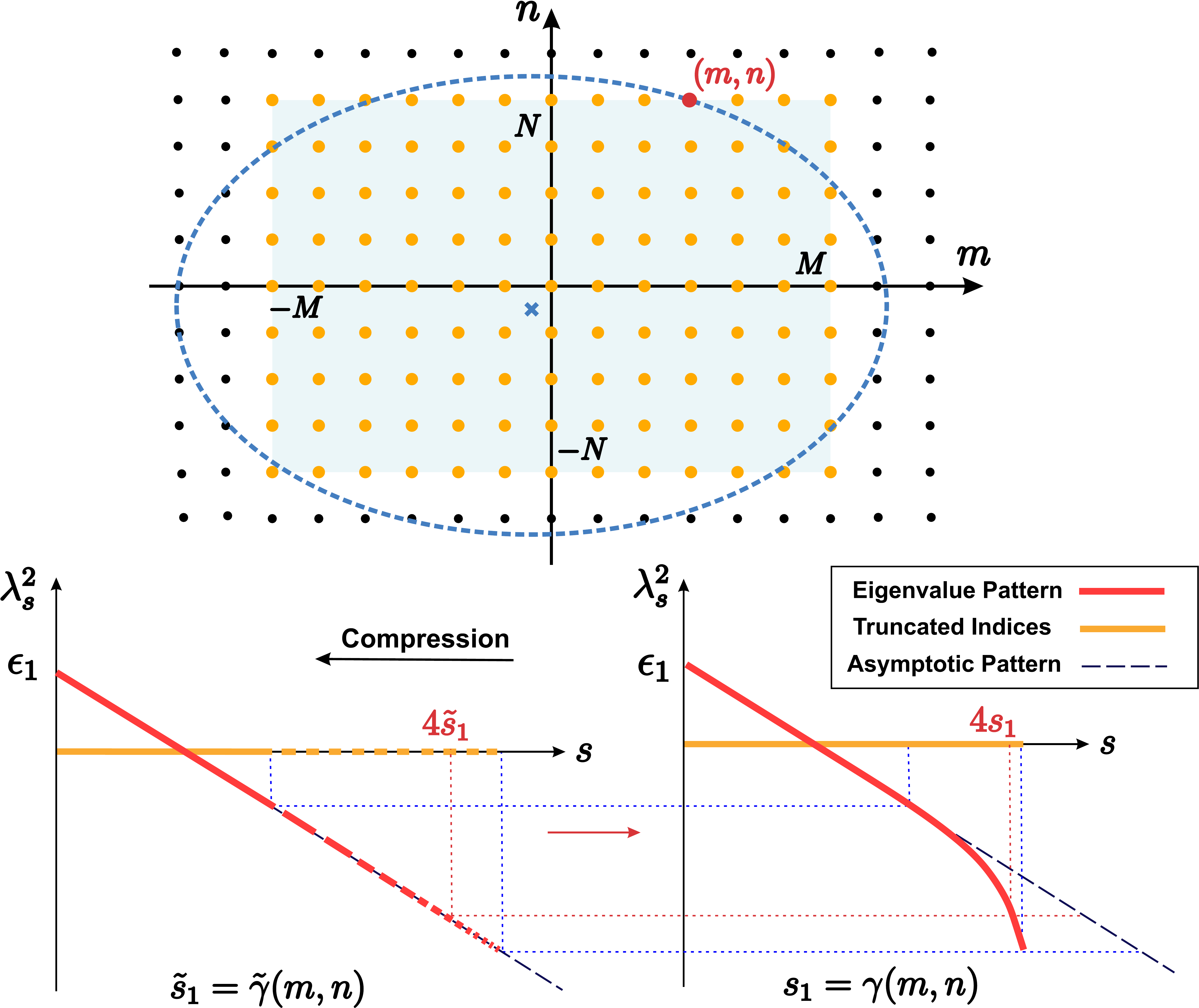}
\caption{Truncation-induced deviation in the eigenvalue pattern of a uniform layer.}
\label{fig_dev}
\end{figure}

In which $\Scale[.8]{{{\rm \tilde g}_{1}}( s ) := {\epsilon _1} - {( {4\pi {T_x}{T_y}})^{ - 1}}s}$ is the asymptotic pattern of a uniform layer with the dielectric constant $\Scale[.95]{\epsilon_1}$. The idea has been schematically depicted in Fig.~\ref{fig_ellip}. Three typical characteristic ellipses are plotted on a truncated integral lattice in Fig.~\ref{fig_ellip_a}. The eigenvalue and the asymptotic patterns are plotted in Fig.~\ref{fig_ellip_b}, with corresponding indices of the three ellipses specified on the $s$-axis.
Let us highlight an important point here, regarding the asymptotic pattern of a grating. It might be inferred from Fig.~\ref{fig_ellip_b} that for a truncated problem, there are always at least a portion of $\Scale[.9]{1-\pi/4 \approx 22\%}$ deviated eigenvalues from the aforementioned asymptotic pattern. Hence, by increasing truncation orders, the maximum error tends to infinity and the word "asymptotic" seems to be misused here. However, the reader should note that this deviation is only a visual byproduct caused by 1D depiction of eigenvalues in a 2D problem.
In fact, because of truncation, some of the more dominant eigenvalues corresponding with lower $s$ indices do not appear in $\Scale[0.9]{ \bm\sigma(\bm Q )}$, while less dominant ones may be present. Consequently, a deviation in the eigenvalue pattern shows up due to omission of unused indices and compression of the horizontal $s$-axis. Nevertheless, they can still be recovered around the asymptote at the cost of getting a less practical illustration, using an order function $\Scale[0.9]{\tilde \gamma:\mathbb{Z}^2 \longleftrightarrow \mathbb{Z}^{+}}$ which has the same description of Eq.~\ref{Eq-14.5}, only defined on the whole lattice points.
This phenomenon has been schematically depicted in Fig.~\ref{fig_dev}, where every lattice point corresponds to an eigenvalue of the structure, while only the yellow ones are obtained due to truncation. The characteristic ellipse of an eigenvalue is depicted. Since the ellipse is large enough to enclose some of the black points as well, the corresponding value lies on the deviated part of the eigenvalue pattern. Another approach to characterize this behavior would be to count the integral points of the truncated lattice (instead of the infinite lattice) lying within the eigenvalue ellipses of Eq.~\ref{Eq-13}, and derive a nonlinear relationship for $\Scale[.85]{s = {{\rm \tilde g_{1}}^{-1}}( \lambda_s^2 )}$. Using either way to characterize the truncation-induced deviation of eigenvalues for uniform layers, can help describe the general case later.

\subsection{General Gratings}
\label{Subsec: Gen}

Now, we consider a general grating, comprised of $\Scale[.95]{L}$ dielectrics $\Scale[0.85]{ {\epsilon _1} > {\epsilon _2} >  \cdots  > {\epsilon _L} }$, with corresponding filling factors $\Scale[0.85]{ {f_1},{f_2}, \ldots ,{f_L} }$ such that $\Scale[0.85]{ \sum\nolimits_{\Scale[.75]{k = 1}}^{\Scale[.75]{L}} {{f_k}}=1 }$. In Eq.~\ref{Eq-9}, the modal matrix $\Scale[.9]{\bm{P}}$ is separated into two characterizing summands, which happens to play an important role in our estimation approach. Despite the fact that only for a uniform grating, the second summand is guaranteed to vanish, numerical evaluations reveal that the spectrum of the modal matrix $\Scale[.9]{\bm P}$ is very little affected by the second summand at all. In other words, the eigenvalue pattern is governed only by the first summand, and following approximation holds in general:

\begin{equation}\label{Eq-16}
\Scale[.95]{
\bm{\mathcal{C}}_{\epsilon} \approx \bm\sigma {(\bm Q )^4}.
}
\end{equation}

On the other hand, $\Scale[.9]{\bm{Q}}$ is the sum of the two matrices $\Scale[0.9]{ {\bm F}}$ and $\Scale[0.85]{ -{{\bm N_x^2}} -{{\bm N_y^2}}}$. Let us consider each of the two summands individually. The matrix $\Scale[0.85]{ -{{\bm N_x^2}} -{{\bm N_y^2}}}$ is a diagonal matrix and its spectrum can easily be characterized using Eq.~\ref{Eq-14} or Eq.~\ref{Eq-15}, by setting $\Scale[0.9]{ \epsilon_1=0}$. As for $\Scale[0.9]{ {\bm{F}}}$, notice the following asymptotic equivalence between the eigenvalue equation in the matrix form and the functional form, in which bounded functions $\Scale[0.9]{\epsilon(x,y)}$ and $\Scale[0.9]{v(x,y)}$ are defined on the unit cell:

\begin{equation}\label{Eq-17}
\Scale[.95]{
( {\bm{F}  - \lambda \bm{I}}){\bm v} = 0 {\hspace{10pt}}\Leftrightarrow {\hspace{10pt}}\left({\epsilon( {x,y}) - \lambda }\right)\times v({x,y}) = 0.
}
\end{equation}

The above equation implies that every eigenvalue of the matrix $\Scale[.95]{\bm{F}}$ at the limiting point is a value $\Scale[0.9]{ \epsilon ({x,y})}$ takes. Therefore, using multi-set notations:

\begin{equation}\label{Eq-18}
\Scale[.9]{
{\bm \sigma}({\bm F}) \approx \mathop  {\displaystyle\bigcup \limits_{k \in {\mathcal{K}_L}} {\left\{ {{\epsilon _k}} \right\}^{\Scale[.7]{{f_k}S}}}}.
}
\end{equation}

Which means that for any $\Scale[.85]{k \in {\mathcal{K}_L}}$, a portion of $\Scale[.9]{{f_k}S}$ out of the total $\Scale[.95]{S}$ eigenvalues of the matrix $\Scale[0.95]{ \bm{F}}$ approximately equal $\Scale[0.95]{ \epsilon_k}$. Now, we'd like to highlight a pivotal relationship between the eigenvalue spectra of $\Scale[0.9]{\bm{Q}}$ and its constituting summands. Numerical evaluations suggest that $\Scale[0.85]{ {\bm\sigma}(\bm{Q})}$ approximately results from element-wise summation of $\Scale[0.8]{ {\bm\sigma}(-{\bm N_x^2}-{\bm N_y^2})}$ to a permutation of $\Scale[0.85]{{\bm\sigma}(\bm{F})}$. In other words, any eigenvalue of $\Scale[0.9]{ \bm{Q}}$ can approximately be expressed as $\Scale[0.85]{ {q _s} = {\epsilon_k} - N_{xm}^2 - N_{yn}^2}$ in which $\Scale[.9]{(m,n)}$ is a unique index pair, and $\Scale[0.9]{ { k \in \mathcal{K}_L}}$ is randomly selected with a probability of $f_k$. In Fig.~\ref{fig_gen_a}, eigenvalue spectra of $\Scale[0.95]{ \bm{Q}}$ and its constituting summands are comparatively depicted. Note that using Eq.~\ref{Eq-18}, a similar property can be proven for the summation of two finite block Toeplitz matrices. An immediate consequence of this property is that the eigenvalue pattern $\Scale[.85]{{\rm g}(s)}$ can be estimated using characteristic patterns of the constituting dielectrics, i.e. $\Scale[0.85]{ {\rm {g_k}}( s )={\rm {g}}( s )|_{{\epsilon (x,y)}={\epsilon_k}}}$. To clarify this relationship, let $\Scale[0.9]{ {q_s}}$ denote the $s$-th element of $\Scale[0.85]{ {\bm\sigma}(\bm{Q})}$ in descending order. Notice that since $\Scale[.9]{q_s \approx \lambda_{4s}^2}$, one can write $\Scale[.85]{s \approx \nicefrac{1}{4}\times {\rm g^{-1}}(\lambda_{4s}^2)}$ for $\Scale[.9]{s \in \mathcal{K}_S}$. We define the lattice point $\Scale[.9]{(m,n)}$ to be $k$-labelled, if we have $\Scale[0.9]{ {q _{s}} = {\epsilon_k} - N_{xm}^2 - N_{yn}^2}$ for some $\Scale[.9]{s \in \mathcal{K}_S}$. Every lattice point is labelled exactly once, and with a corresponding probability of $\Scale[.95]{f_k}$ for $\Scale[.9]{k \in \mathcal{K}_L}$. Now for each eigenvalue $\Scale[.9]{q_s}$, consider $\Scale[.95]{L}$ characteristic ellipses in the $mn$-plane, concentric at $\Scale[0.75]{\left( {-N_{x0}{T_x}},{-N_{y0}{T_y}} \right)}$, and with semi-minor and semi-major axes
$\Scale[0.85]{ r_x^k={T_x}({ {\epsilon_k} - q_s})^{\nicefrac{1}{2}}}$ and $\Scale[0.85]{ r_y^k={T_y}({ {\epsilon_k} - q_s})^{\nicefrac{1}{2}} }$, correspondingly. From previous discussions, the number of lattice points within the $k$-th ellipse is $\Scale[0.85]{s_k =\nicefrac{1}{4}\times {{\rm g^{-1}_k}(\lambda_{4s}^2)}}$. Since the label distribution is random, the number of $k$-labelled lattice points enclosed by the $k$-th ellipse is approximately $\Scale[.9]{{f_k}{s_k}}$. On the other hand, an eigenvalue $\Scale[0.95]{ q_{s'}}$ appear before $\Scale[0.95]{ q_s}$ in the order, i.e. $\Scale[.95]{s'<s}$, if and only if the $k$-th characteristic ellipse of $\Scale[0.95]{q_s}$ encompasses the $k$-th characteristic ellipse of $\Scale[0.95]{q_{s'}}$, for some $\Scale[.85]{k \in \mathcal{K}_L}$ (the center of ellipses can be moved infinitesimally to guarantee no two lattice points lie on the same ellipse). Equivalently, the ordinal index $s$ equals the number of $k$-labelled points in the $k$-th ellipse added together, i.e. $\Scale[.9]{s \approx \sum\nolimits_{\Scale[.73]{k=1}}^{\Scale[.7]{L}} {\Scale[.95]{{f_k}{s_k}}}}$. Rewriting both sides of the equation in terms of $\Scale[.85]{\rm g^{-1}}$-functions and by changing $\Scale[.9]{4s \to s}$, a simple and beautiful relationship is obtained as following:

\begin{equation}\label{Eq-19}
\Scale[.95]{
{\rm g}^{-1}(\lambda_s^2)\approx {\displaystyle\sum\nolimits_{k=1}^{L} {{f_k}{\rm g}_k^{-1}}(\lambda_s^2)}.
}
\end{equation}

The above equation expands the inverse eigenvalue pattern of a crossed grating as a weighted summation over inverse eigenvalue patterns of the constitutive dielectrics. In Fig.~\ref{fig_gen_b}, a schematic description of the rule of average eigenvalue patterns is presented. In order to provide a better insight into the asymptotic behavior of the eigenvalues and to get a simpler estimation, one may utilize Eq.~\ref{Eq-15} to approximate the eigenvalue patterns of uniform layers. It will be straightforward then to extract the asymptotic pattern of a grating as a piece-wise linear function, in terms of the structural parameters:

\begin{figure}[!t]
\centering
	\begin{subfigure}{0.5\textwidth}
		\centering
		\includegraphics[width=0.85\linewidth]{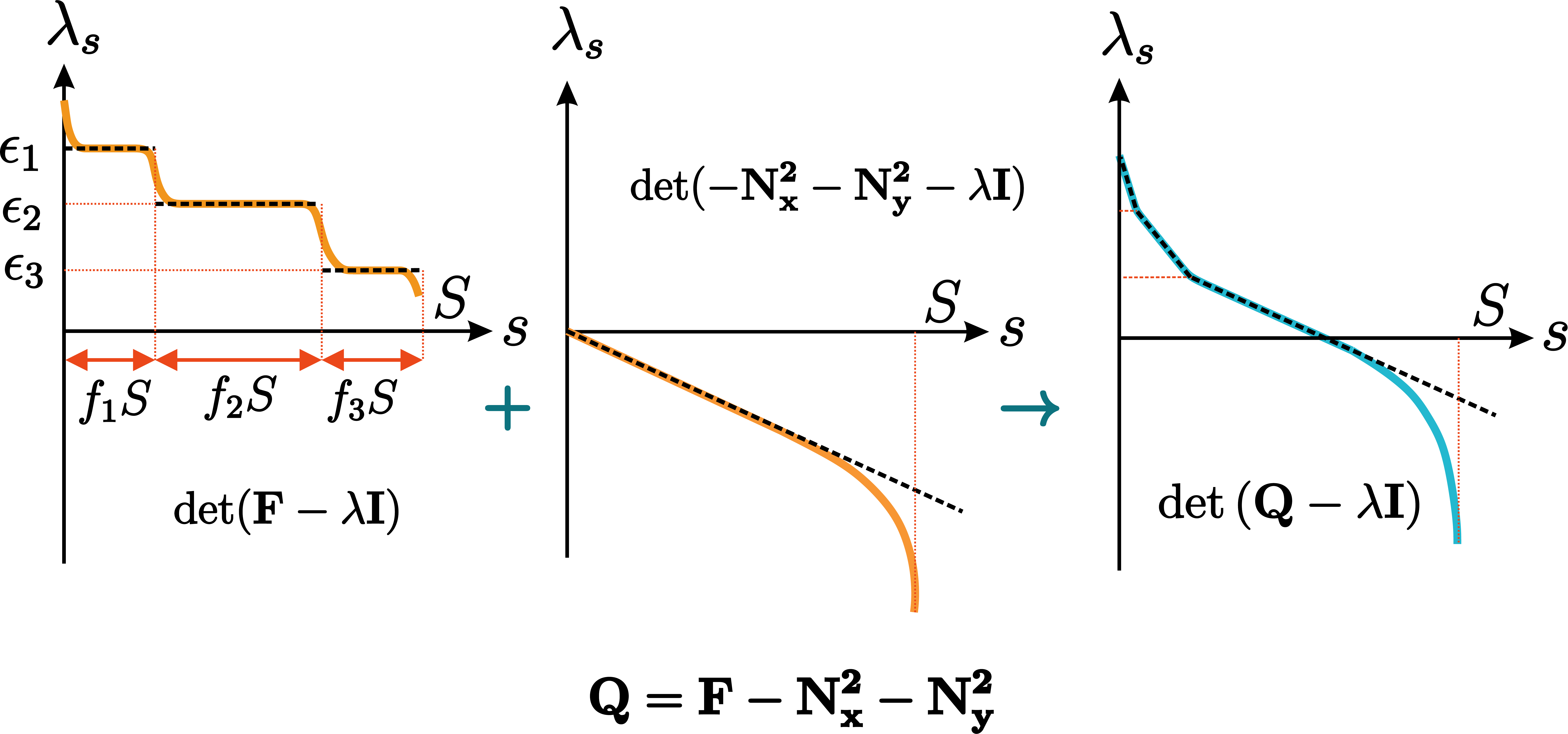}
		\caption{}\label{fig_gen_a}		
	\end{subfigure}
\quad
	\begin{subfigure}{0.5\textwidth}
		\centering
		\includegraphics[width=0.84\linewidth]{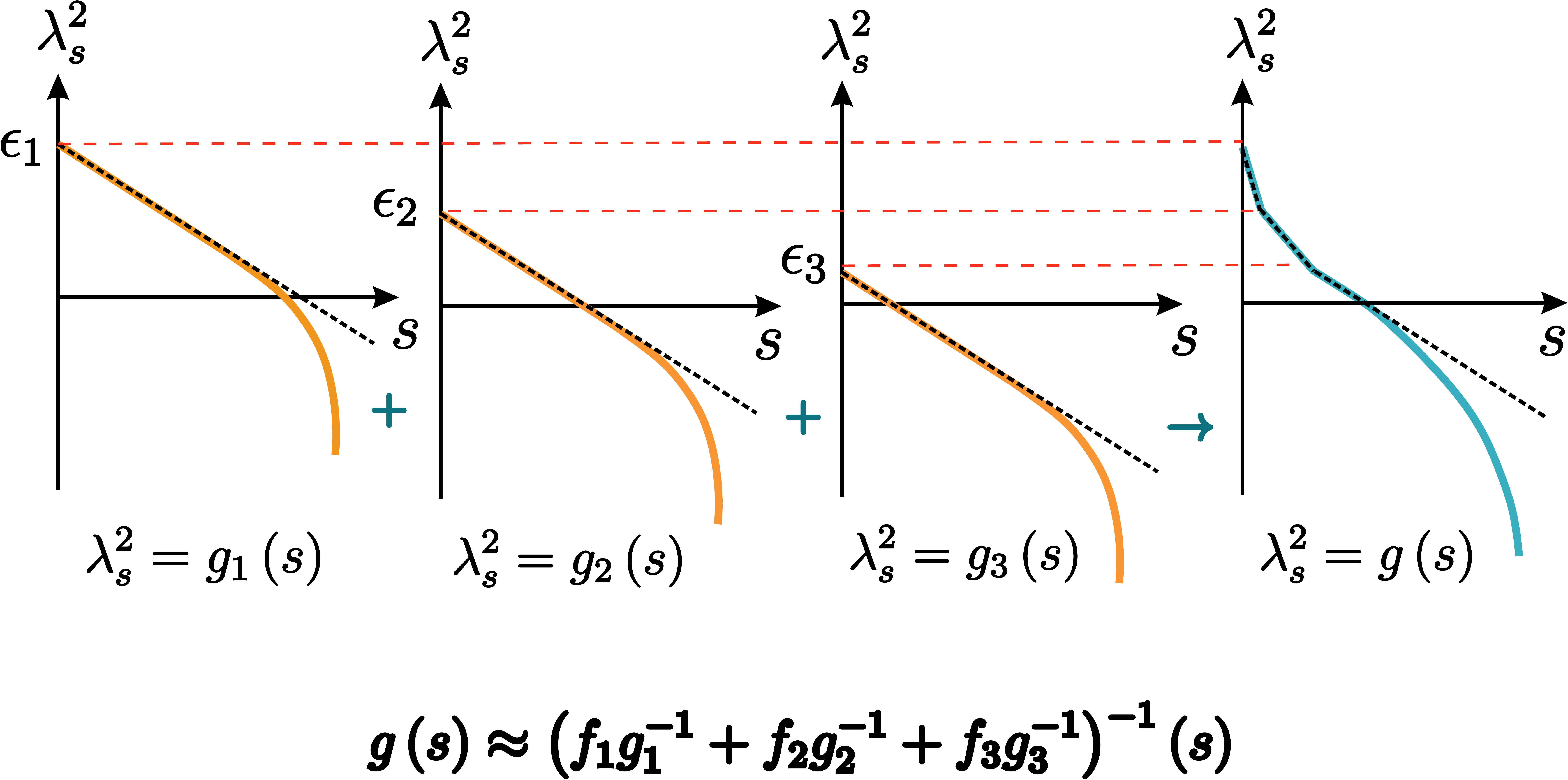}
		\caption{}\label{fig_gen_b}		
	\end{subfigure}
\quad
	\begin{subfigure}{0.5\textwidth}
		\centering
		\includegraphics[width=0.75\linewidth]{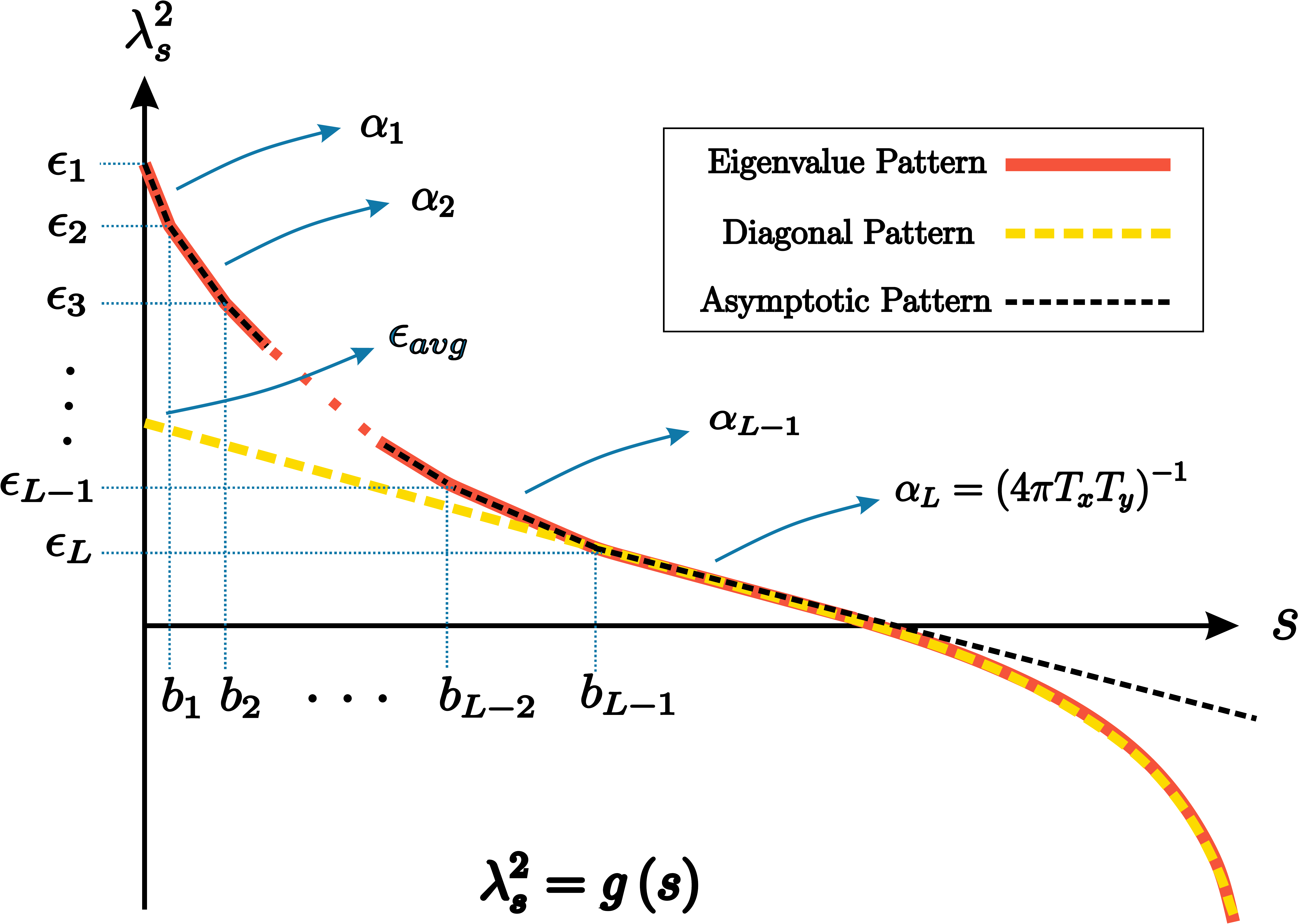}
		\caption{}\label{fig_gen_c}
	\end{subfigure}

\caption{(\subref{fig_gen_a}) The eigenvalue patterns of $\Scale[.9]{\bm Q}$ and its two constituting summands. (\subref{fig_gen_b}) The eigenvalue pattern of a grating is constructed through weighted summation of characteristic patterns (obtained from uniform layers) along the horizontal $s$-axis. (\subref{fig_gen_c}) The eigenvalue pattern, the diagonal pattern, and the asymptotic pattern for the generalized grating problem consisting of $L$ dielectrics.}
\label{fig_gen}
\end{figure}

\begin{eqnarray}\label{Eq-20}
\begin{array}{l}
\Scale[0.89]{\hspace{5pt}{\rm {\tilde g}}( s ) = {\epsilon _k} - {\alpha _k}( {s - {b_{k - 1}}} )\hspace{51pt} ({b_{k - 1}} < s \le {b_k} )}.\\
\Scale[0.84]{\left\{ \begin{array}{l}
\alpha _k^{ - 1} = ( {4\pi {T_x}{T_y}} ){\displaystyle\sum\nolimits_{i = 1}^k {{f_i}}\hspace{54pt}({k \in \mathcal{K}_L})} \\
{b_k} = {\displaystyle\sum\nolimits_{i = 1}^k {\alpha _i^{ - 1}( {{\epsilon _{i + 1}} - {\epsilon _i}} )}\hspace{51pt}({k \in \mathcal{K}_{L-1}})}
\end{array} \right.}.
\end{array}
\end{eqnarray}

In which $\Scale[0.85]{ 0 = {b_0} < {b_k} <  \cdots < {b_L} = \infty }$ are the break points of the pattern on the horizontal axis, and $\Scale[0.85]{ {\alpha _1} > {\alpha _k} >  \cdots  > {\alpha _L}}$ are the slopes of the linear segments without the negative sign, for $\Scale[0.85]{ k \in \mathcal{K}_L}$. The eigenvalue pattern and the asymptotic pattern of a crossed grating are schematically depicted in Fig.~\ref{fig_gen_c}. In this figure, notice that after the last break point on the $s$-axis, i.e. $\Scale[.85]{s > b_{L-1}}$, the eigenvalue pattern is quite smooth and linear. It can simply be shown that $\Scale[.85]{b_{L-1}=4\pi{T_x}{T_y}(\epsilon_{avg} - \epsilon_{L})}$, with $\Scale[.9]{\epsilon_{avg} \approx \sum\nolimits_{\Scale[.7]{k=1}}^{\Scale[.7]{L}} {\Scale[.93]{{f_k}{\epsilon_k}}}}$ as the average dielectric constant of the grating. Moreover, the equation of the last linear segment of the pattern will be $\Scale[0.85]{\lambda_s^2 = {\epsilon_{avg}} - {( {4\pi {T_x}{T_y}} )^{-1}}s}$, which is the same as the asymptotic pattern of a uniform layer with the dielectric constant $\Scale[.85]{\epsilon_{avg}}$, and also implies that the number of propagating modes approximately equals to $\Scale[.85]{4\pi {T_x}{T_y}{\epsilon _{avg}}}$, for large  enough truncation orders.

\subsection{Diagonal Patterns}
\label{Subsec: Diag}

Notwithstanding that Eq.~\ref{Eq-19} and Eq.~\ref{Eq-20} uncover a fascinating property of gratings and provide insightful approximations for propagation constants, the estimation can partly deteriorate at initial points due to oversimplification, when complicated shapes or high dielectric contrasts are involved. Although this approximation error hardly materializes in real problems, it's still worth finding a practical solution to deal with it. Here, we'd like to provide a technique for  estimation of the propagation constants, based on the diagonal entries of the modal matrix $\Scale[0.95]{ \bm{P}}$. To clarify this idea, define the multi-set $\Scale[.85]{\bm{\mathcal{D}}_{\epsilon} ={\bm\delta} {( \bm P )^2}}$, sort its elements in descending order, and let the $s$-th element be denoted by $\Scale[.9]{{\rm{ d}}( s )}$. Note that the diagonal entries of $\Scale[0.95]{ \bm{P}}$ are real numbers. If we plot $\Scale[.9]{{\rm d}{(s)}}$ against the sequential index $\Scale[.85]{s \in {\mathcal{K}_{4S}}}$, a particular pattern is obtained which will be named as the "diagonal pattern". It can be shown that the diagonal pattern of a grating approximately equals to the eigenvalue pattern of a uniform layer with an average dielectric constant, i.e. $\Scale[.85]{{\rm d}(s) \approx \left.{\rm g}(s)\right|_{\Scale[.8]{\epsilon = \epsilon_{avg}}}}$ (see the appendix).
As a result, the linear part of $\Scale[.9]{{\rm d}(s)}$ coincides with the last linear segment of $\Scale[.9]{{\rm {\tilde g}}(s)}$, as schematically depicted in Fig.~\ref{fig_gen_c} for a general grating. Moreover, the inherent deviation of the diagonal pattern practically pursues the eigenvalue pattern. This surprising result means that all but the first $\Scale[0.9]{ b_{L-1}}$ propagation constants appear on the main diagonal of the modal matrix $\Scale[.95]{\bm P}$, and can immediately be extracted without solving an equation. The reader should note that the diagonal property of the modal matrix $\Scale[.99]{\bm{P}}$ typically can not be inferred from the Gershgorin circle theorem~\cite{Brualdi1994, varga2004gresch}, as radii of corresponding circles become uninformatively large and the circles intersect. On the other hand, a principal property of the eigenvalue pattern which makes its definition practically useful is that by increasing truncation orders, not only does $\Scale[0.85]{{\rm {g}}(s )={\rm Re}({\lambda _s^2})}$ characterize a well-defined asymptotic pattern, but also it fills out the pattern in the same order as the eigenvalues were sorted. More precisely, for large enough arbitrary truncation orders $\Scale[.85]{(M_1,N_1)}$ and $\Scale[.85]{(M_2,N_2)}$ with corresponding total orders $\Scale[.85]{{S_1}<{S_2}}$, provided that $\Scale[.85]{{M}/{N}=\rho}$ remains constant, there exists $\Scale[.85]{0 < C_\rho < 1}$ such that:

\begin{equation}\label{Eq-21}
\Scale[.95]{
{\left. {\rm g(s)} \right|_{S = {S_1}}} \approx {\left. {\rm g(s)} \right|_{S = {S_2}}},\hspace{45pt}(\hspace{2pt}s \in {\mathcal{K}_{4{C_\rho}{S_1}}}).
}
\end{equation}

It simply means that after a certain order, eigenvalue patterns corresponding to different truncation orders match at the points with lowest ordinal indices. For example, one can easily show that $\Scale[.85]{C_{\rho} =\nicefrac{\pi}{4}}$ for a uniform layer, provided that $\Scale[.85]{\rho={T_x}/{T_y}}$. An illustration of this idea has been provided in Fig.~\ref{fig_par_a}, which will be explained in the next section. As a result, in order to estimate the first $\Scale[.9]{b_{L-1}}$ propagation constants of the grating, it suffices to form the eigenvalue equation with an order $\Scale[.9]{S_0}$ and find the first $\Scale[.9]{b_{L-1}}$ propagation constants of it, provided that still $\Scale[.85]{{b_{L-1}} \ll 4{S_0}}$. Combining the ideas, a new scheme can be proposed according to the following approximation:

\begin{equation}\label{Eq-22}
\Scale[.9]{
{\bm{\mathcal{C}}}_\epsilon \approx
\{ \left. {{\bm{\mathcal{C}}}_\epsilon}\right|_{S=S_0} \}_{{}_{\Scale[.7]{b_{L-1}}}} ^ {{}^{\Scale[.7]{\geqslant}}} \cup \{ {{\bm{\mathcal{D}}}_\epsilon} \}_{{}_{\Scale[.7]{4S - {b_{L-1}}}}} ^ {{}^{\Scale[.7] {\leqslant}}},\hspace{18pt}(\Scale[.9]{{b_{L-1}} \ll 4{S_0} \ll 4{S}}).
}
\end{equation}

Wherein by definition, $\Scale[.85]{\lambda_s^2 = \left\langle \bm{\mathcal{C}}_{\epsilon} \right\rangle_{s}}$ provides the estimation. Although this technique still requires solving an eigenvalue equation, it can effectively reduce the computational cost of the initial problem. As in practice, it suffices to take $\Scale[.9]{S_0}$ equal to or a few times $\Scale[.9]{b_{L-1}}$, the size of the second equation can remain independent from $\Scale[.95]{S}$, while the computational cost of the initial problem grows polynomially with $\Scale[.9]{O(S^3)}$. Moreover, if $\Scale[.9]{b_{L-1}}$ becomes negligibly small, the diagonal pattern approximately matches the eigenvalue pattern, i.e. all squared propagation constants appear on the main diagonal of the modal matrix $\Scale[.95]{\bm P}$:

\begin{equation}\label{Eq-23}
\Scale[.95]{
{\lambda _s^2} \approx {\rm d}(s),\hspace{65pt}(\Scale[.85]{{b_{L-1}} \to 0}).
}
\end{equation}

Practically, a $\Scale[.85]{{b_{L-1}} \approx 1}$ sounds ideal for the above estimation to hold, though larger values can still work well, at the cost of a plausible deterioration. This condition covers a lot of practical cases, in which regularly small dielectric constants are used or the contrast is not high. Even for medium contrasts with small enough unit cells, the approximation still holds. Some examples will be provided in the next section.

\section{Numerical Illustration}
\label{Section: Num}

In this section, we will highlight the main aspects of the proposed estimation schemes through a select few of countlessly many possible numerical examples. We show the effect of truncation orders and structural parameters on the eigenvalue pattern, compare the results with the asymptotic patterns, and consider using the diagonal patterns as well. Dielectric constants are mostly chosen unrealistically large, to take the worst case scenario in terms of convergence into account, and to accentuate the segmented structure of the eigenvalue patterns. Nonetheless, examples with realistic dielectric constants are also provided.

\begin{figure}[!ht]		
\centering	\includegraphics[width=0.9\linewidth]{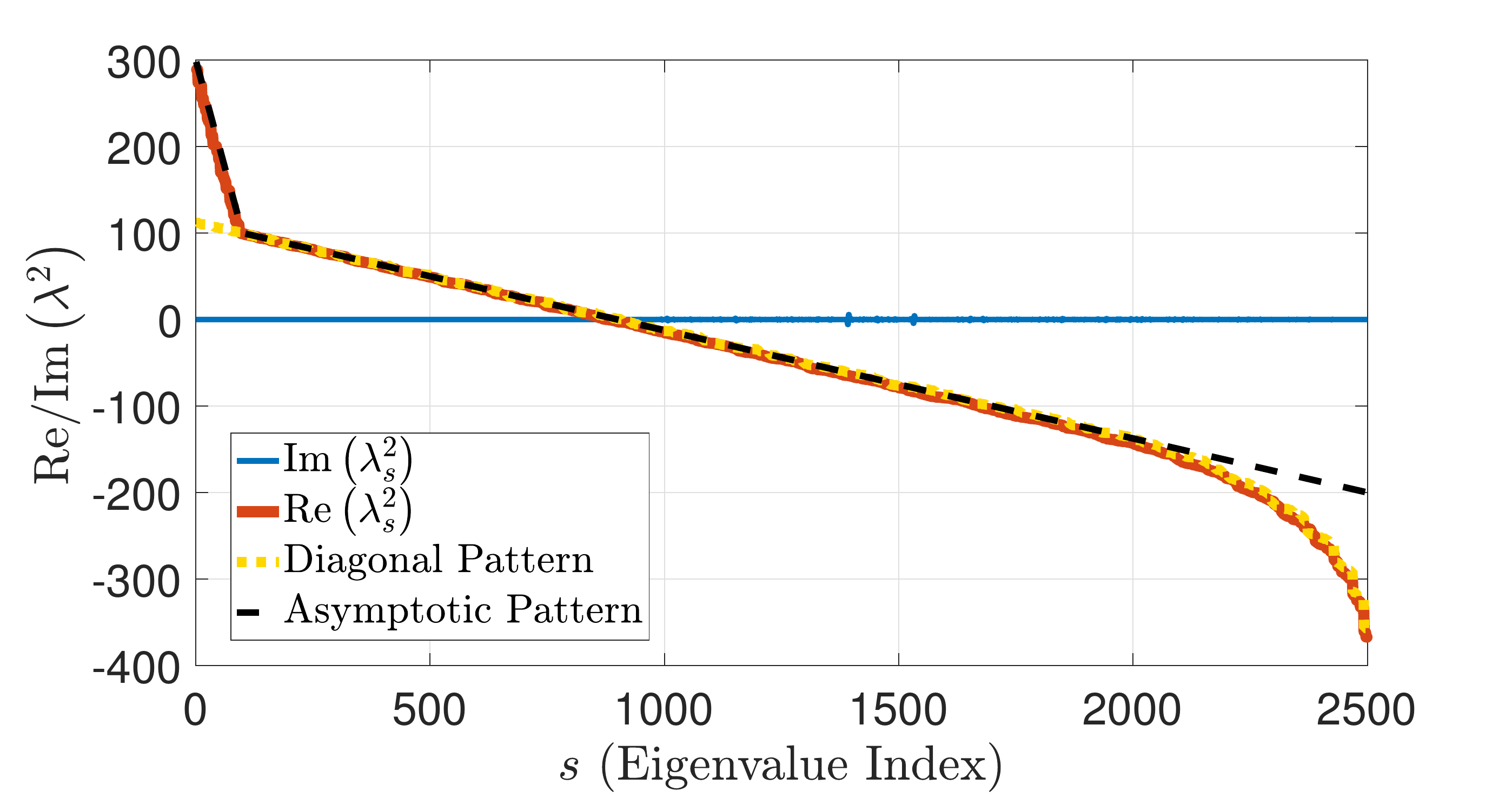}

\caption{The graphs of $\Scale[.85]{{\rm g}{(s)}}$, $\Scale[.85]{{\rm g^*}(s)}$, $\Scale[.85]{{\rm{\tilde g}}{(s)}}$, and $\Scale[.85]{{\rm d}{(s )}}$ for a grating problem with rectangular resonators, $\Scale[0.85]{( {{\epsilon _1},{\epsilon _2}}) = ({300,100})}$, $\Scale[0.85] {{T_x} = {T_y} = 0.8}$, $\Scale[0.85]{ {c_x} = {c_y} = 0.25}$, $\Scale[0.85]{ \theta  = 45^\circ }$, $\Scale[0.85]{ \varphi  = 0^\circ }$ and $\Scale[0.85]{ M = N = 12}$.}

\label{fig_patt}
\end{figure}

\subsection{Depiction of Patterns}
\label{Subsec: Depic}

Consider a grating of rectangular resonators as depicted in Fig.~\ref{fig_3d}, with  normalized unit cell dimensions $\Scale[0.9]{ T_x=T_y= 0.8}$, and resonator dimensions $\Scale[0.9]{ t_x={T_x}/{4}}$ and $\Scale[0.9]{t_y = {T_y}/{4}}$. Dielectric constants $\Scale[0.9]{ \epsilon_1=300}$ for the resonator and $\Scale[0.9]{ \epsilon_2=100}$ for the background dielectric are assumed. Correspondingly, the filling factors will be $\Scale[0.85]{ f_1=\nicefrac{1}{16}}$ for the resonator, and $\Scale[0.9]{ f_2=1-f_1}$ for the background. Let a plane wave be incident upon the grating, with an angle of incidence $\Scale[0.9]{ \theta=45^\circ}$, and an azimuthal angle $\Scale[0.9]{ \varphi=0^\circ}$. Note that eigenvalues are mathematically independent from the polarization angle $\Scale[.9]{\psi}$. The surrounding environment is assumed to be vacuum. Given the specifications with truncation orders $\Scale[0.9]{ M=N=12}$, the set of normalized propagation constants, i.e. $\Scale[0.9]{ {\bm\sigma _2}(\bm P)}$, consists of $\Scale[0.9]{ 4S=2500}$ values. In Fig.~\ref{fig_patt} we have plotted and compared the functions $\Scale[.9]{{\rm {g}}(s)}$, $\Scale[.9]{{\rm {g^*}}(s)}$, $\Scale[.85]{{\rm {\tilde g}}(s)}$, and $\Scale[.85]{{\rm{d}}(s)}$, against the sequential index $s$. It can be observed that the eigenvalue pattern closely collides with the asymptotic pattern, i.e. $\Scale[0.85]{ {\rm {g}( s)}\approx {\rm{\tilde g}( s)}}$, except at the deviated tail of $\Scale[0.9]{{\rm {g}}( s)}$. On the other hand, the diagonal pattern seems to lie on the eigenvalue pattern, i.e. $\Scale[.9]{{\rm {g}(s)}\approx{\rm {d}}(s)}$, though only after the breakpoint of $\Scale[0.9]{{\rm {g}}( s)}$.

\begin{figure}[!ht]	
	\centering

	\begin{subfigure}{0.5\textwidth}
		\centering
		\includegraphics[width=0.84\linewidth]{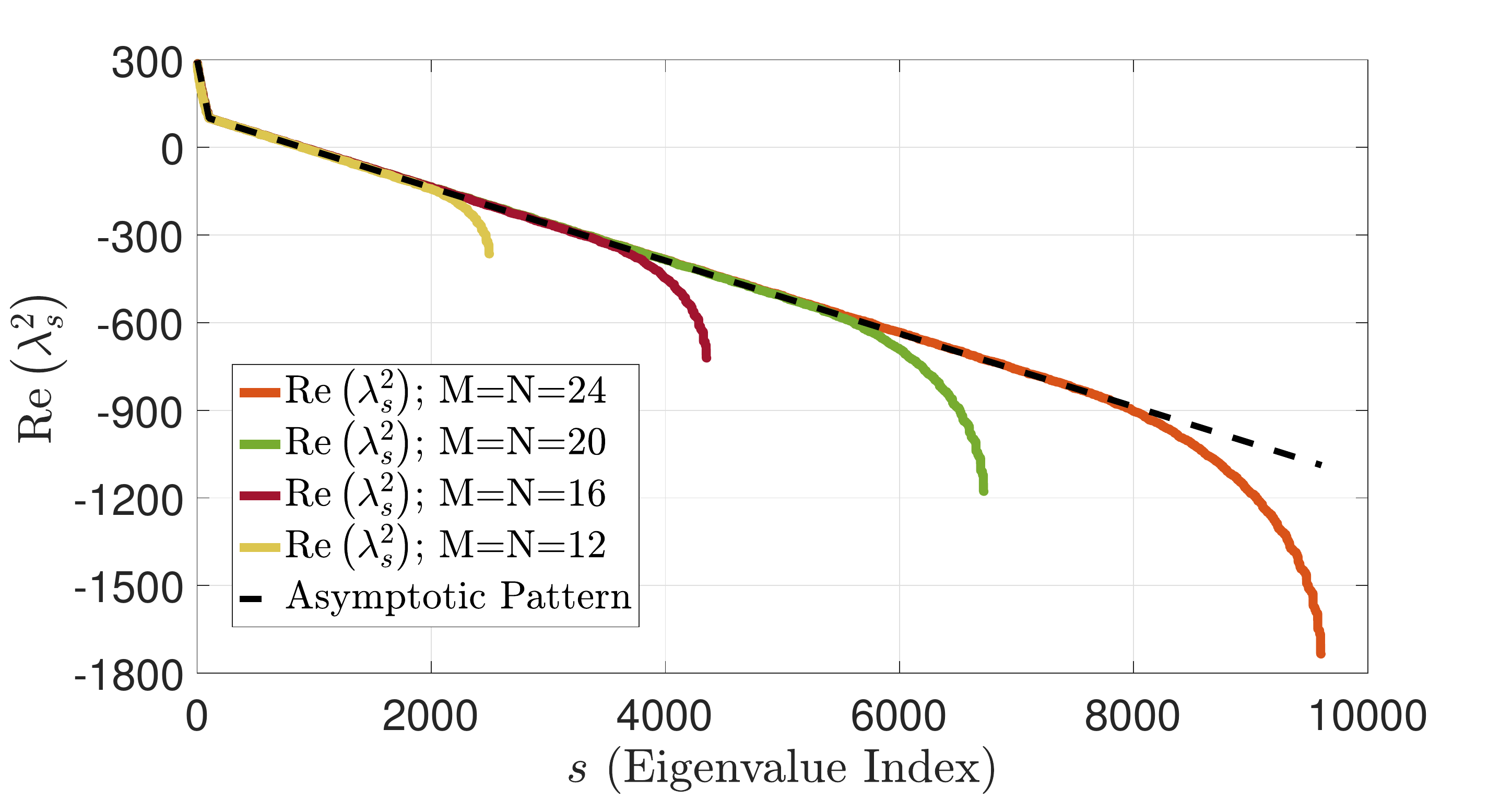}
		\caption{}\label{fig_par_a}
	\end{subfigure}
~
	\begin{subfigure}{0.5\textwidth}
		\centering
		\includegraphics[width=.84\linewidth]{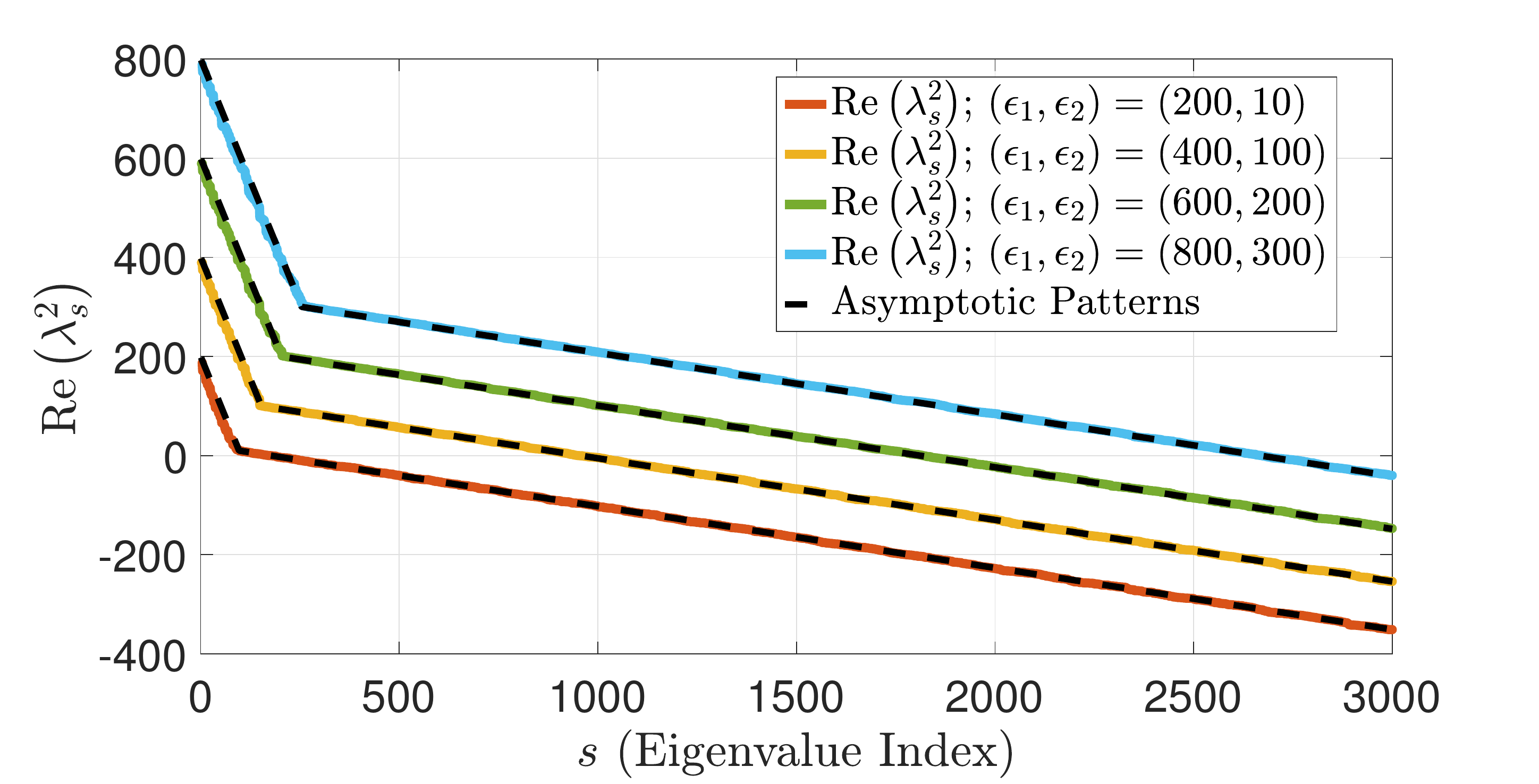}
		\caption{}\label{fig_par_b}
	\end{subfigure}	
~
	\begin{subfigure}{0.5\textwidth}
		\centering
		\includegraphics[width=.84\linewidth]{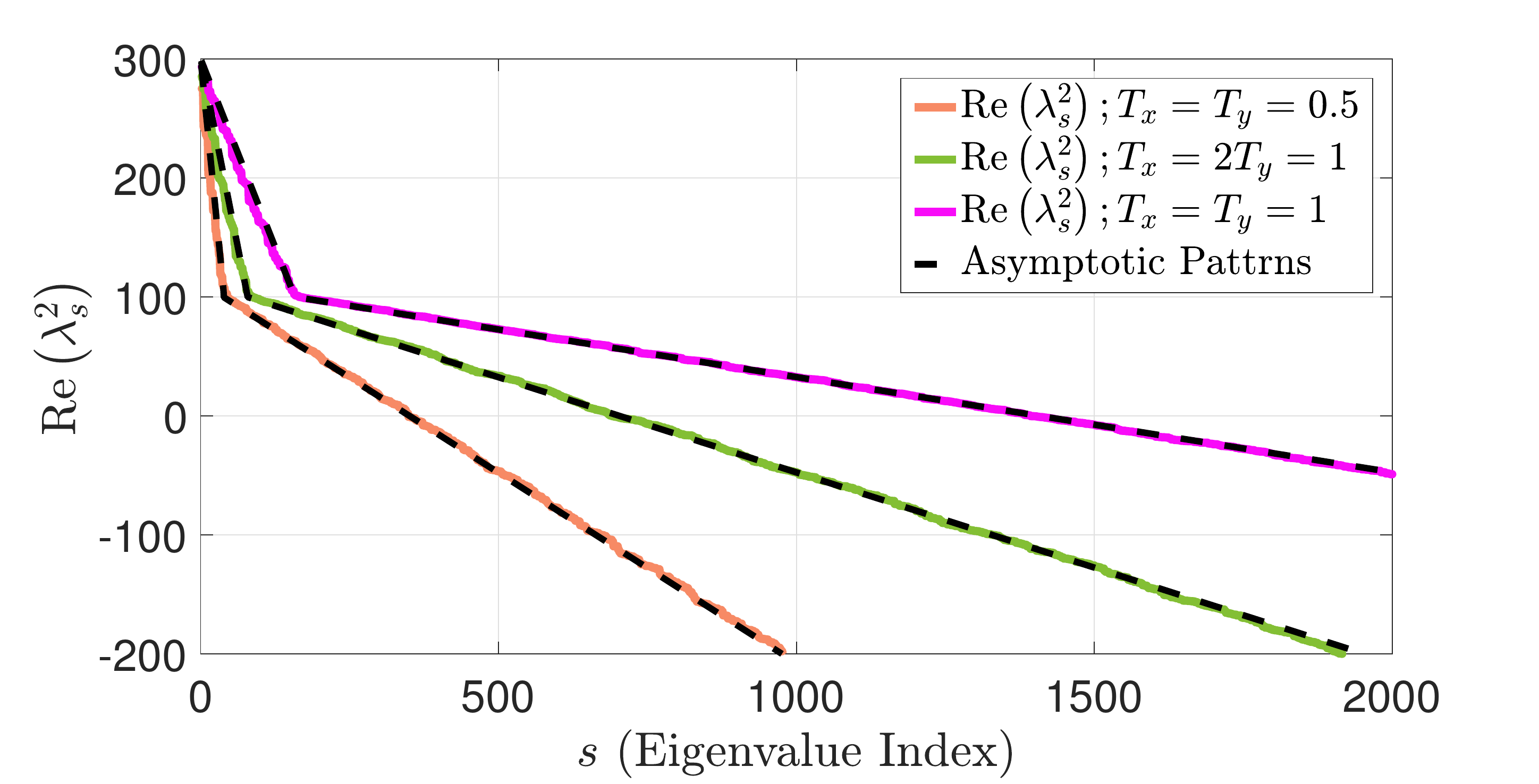}
		\caption{}\label{fig_par_c}
	\end{subfigure}
~
	\begin{subfigure}{0.5\textwidth}
		\centering
		\includegraphics[width=.84\linewidth]{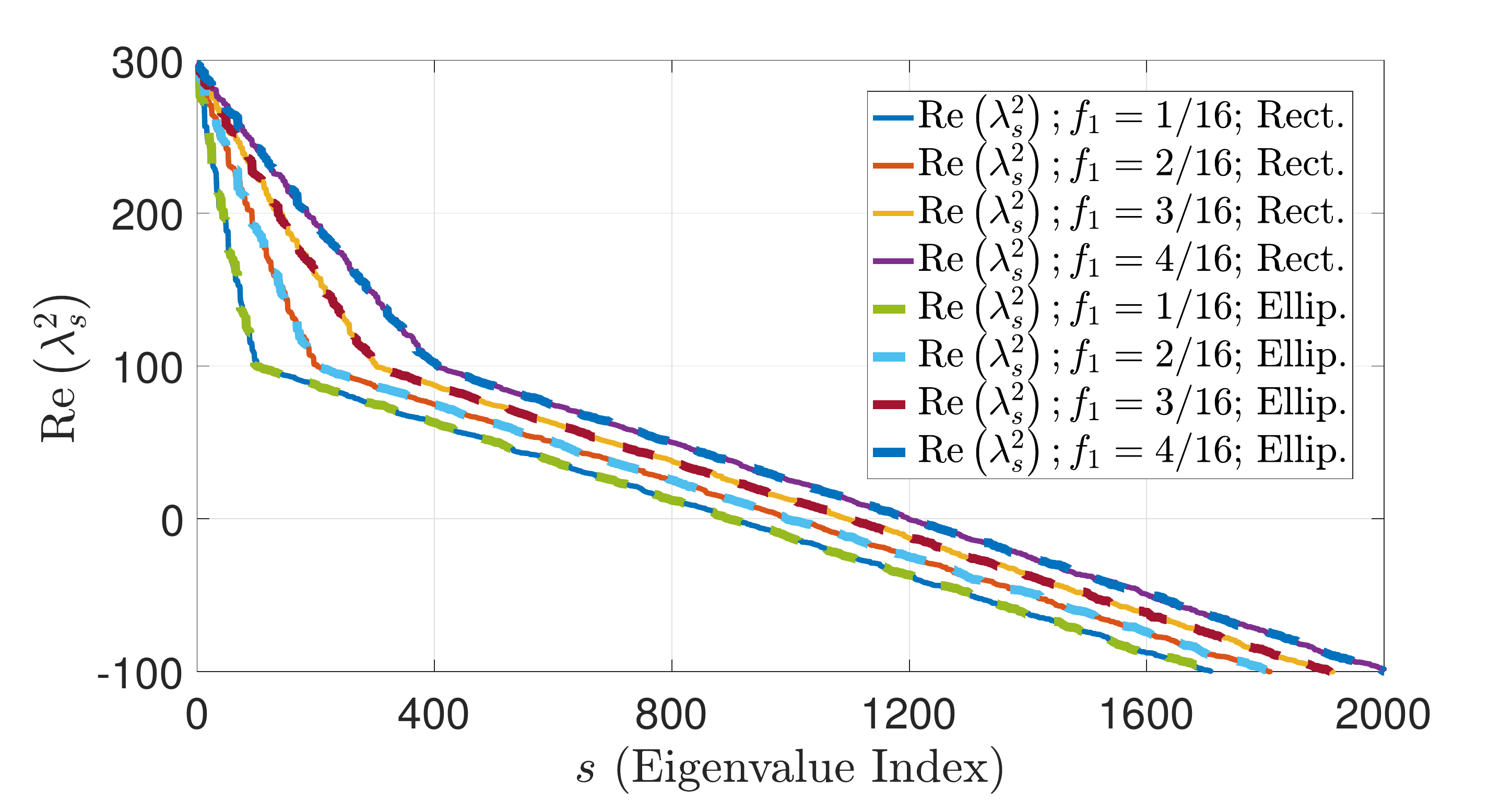}
		\caption{}\label{fig_par_d}
	\end{subfigure}

	\caption{The effect of changing different parameters on the eigenvalue pattern in grating example of Sub.\ref{Subsec: Gen} with,(\subref{fig_par_a} different truncation orders; (\subref{fig_par_b}) different dielectric constants; (\subref{fig_par_c}) different unit cell dimensions; (\subref{fig_par_d}) different filling factors and resonator shapes (rectangular and elliptical).}
	\label{fig_param}
\end{figure}

\subsection{Change of Parameters}
\label{Subsec: Param}

In this part, we have demonstrated the effect of structural parameters and truncation orders on the eigenvalue pattern of a typical grating. To this end, the grating example of Sub.~\ref{Subsec: Depic} is considered and the effect of varying individual parameters on its eigenvalue pattern is explored. We expect the eigenvalues to pursue an asymptotic pattern, as described by Eq.~\ref{Eq-20}. Note that besides the grating structure, eigenvalues also pertain to the parameters of the incident wave and the definition of the order function $\Scale[.9]{\nu (m,n)}$, however, having negligible effect, we have omitted them for brevity. We start by plotting $\Scale[0.9]{{\rm {g}}(s)}$ and $\Scale[0.9]{{\rm {\tilde g}}(s)}$ for different truncation orders $\Scale[0.9]{M =N=12}$, $\Scale[0.9]{16}$, $\Scale[0.9]{20}$, and $\Scale[0.9]{24}$, in Fig.~\ref{fig_par_a}. The sequential completion of an asymptotic pattern, as described by Eq.~\ref{Eq-21}, is totally clear. Recall from Sub.~\ref{Subsec: Unif} that the truncation-induced deviation at the tail of the eigenvalue pattern was just an illustrative implication, and hence did not contradict with the existence of an asymptotic pattern. On the other hand, according to Eq.~\ref{Eq-20}, $\Scale[0.85]{{\rm {\tilde g}}(s)}$ can be specified solely by dielectric constants, cell dimensions and filling factors. In Fig.~\ref{fig_par_b}, we have plotted $\Scale[0.9]{{\rm {g}}(s)}$ and $\Scale[0.9]{{\rm {\tilde g}}(s)}$ for dielectric constants $\Scale[0.9]{ (\epsilon_1,\epsilon_2) =(200,10)}$, $\Scale[0.9]{(400,100)}$, $\Scale[0.9]{(600,200)}$, and $\Scale[0.9]{(800,300)}$. In Fig.~\ref{fig_par_c}, same plots are produced for cell dimensions $\Scale[0.9]{ T_x=T_y=0.5}$, $\Scale[0.9]{ T_x=2T_y=1}$, and $\Scale[0.9]{ T_x=T_y=1}$. Finally in Fig.~\ref{fig_par_d}, the resonator is scaled and $\Scale[0.9]{{\rm {g}}(s)}$ is plotted for filling factors $\Scale[0.9]{ f_1=\nicefrac{1}{16}}$, $\Scale[0.9]{\nicefrac{2}{16}}$, $\Scale[.9]{\nicefrac{3}{16}}$, and $\Scale[0.9]{ \nicefrac{4}{16}}$, with both rectangular resonators (solid lines) and equivalent elliptical ones (dashed lines). Comparing the patterns, asymptotic independence of the eigenvalues from the resonator shape can simply be inferred.

\begin{figure}[!ht]	
	\centering

\begin{subfigure}[!t]{0.13\textwidth}
		\centering
		\includegraphics[width=\linewidth]{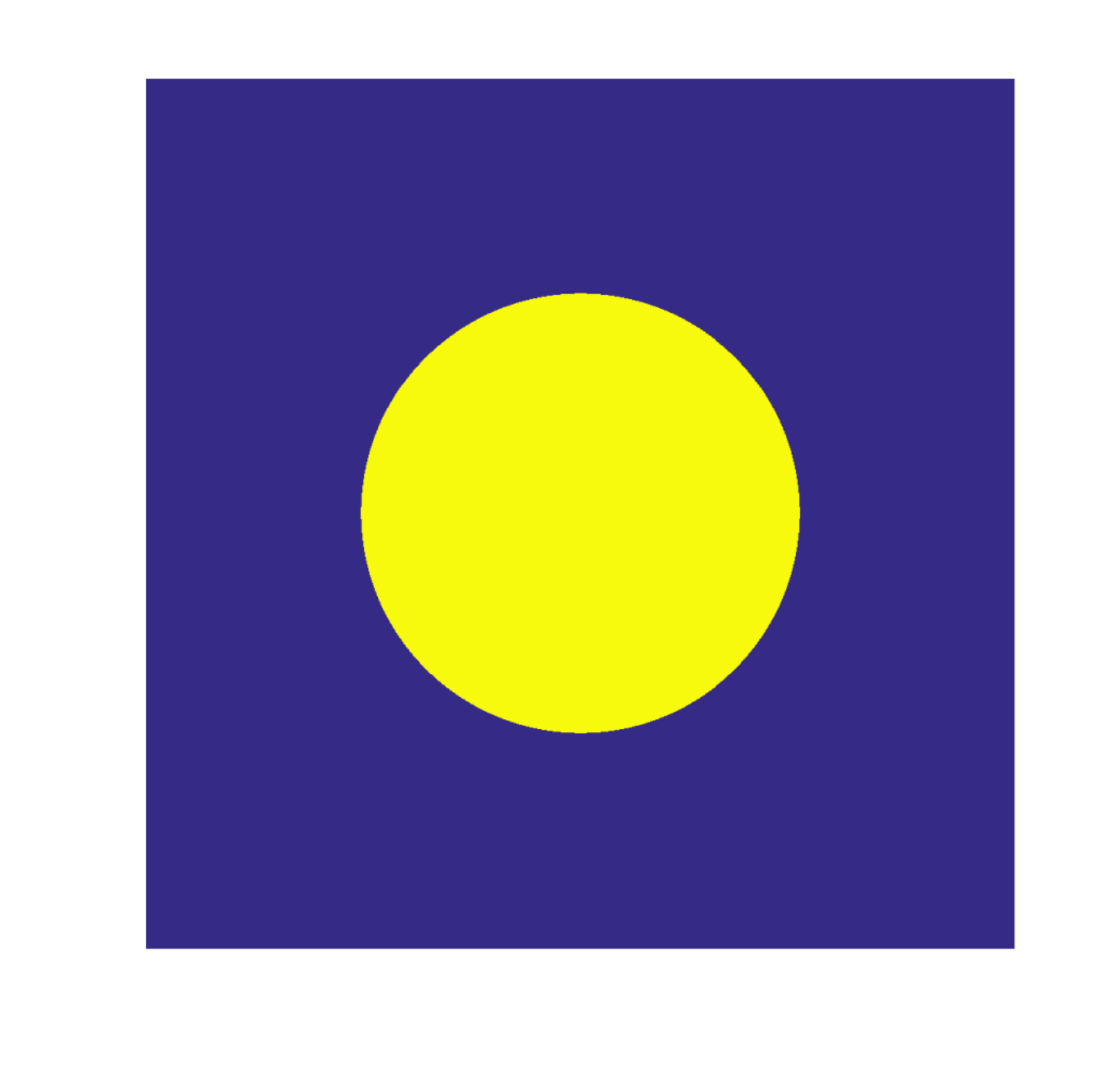}
		\caption{}\label{fig_cplx_a}		
	\end{subfigure}
~
	\begin{subfigure}[!t]{0.13\textwidth}
		\centering
		\includegraphics[width=\linewidth]{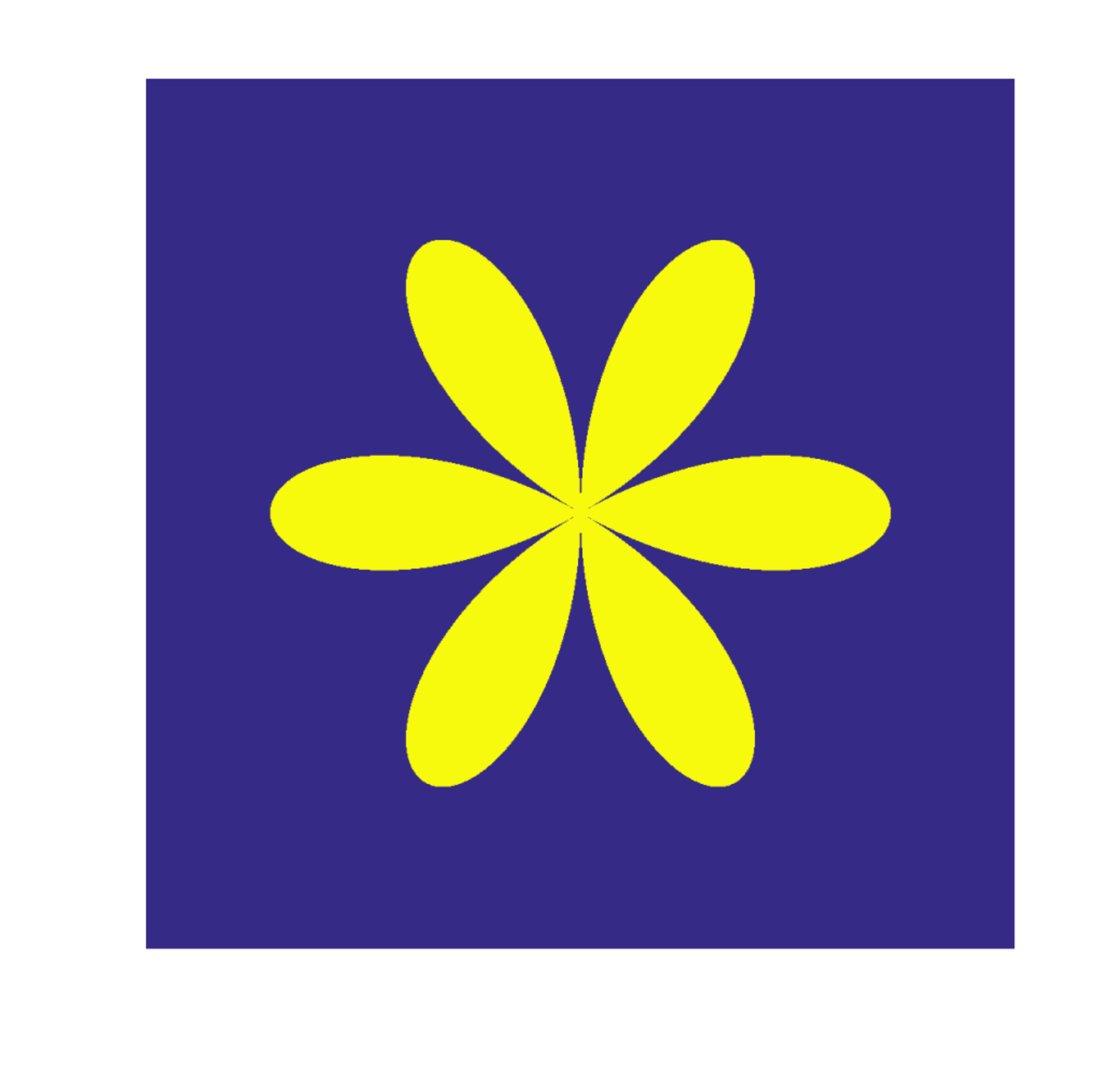}
		\caption{}\label{fig_cplx_b}
	\end{subfigure}
~
	\begin{subfigure}[!t]{0.13\textwidth}
		\centering
		\includegraphics[width=\linewidth]{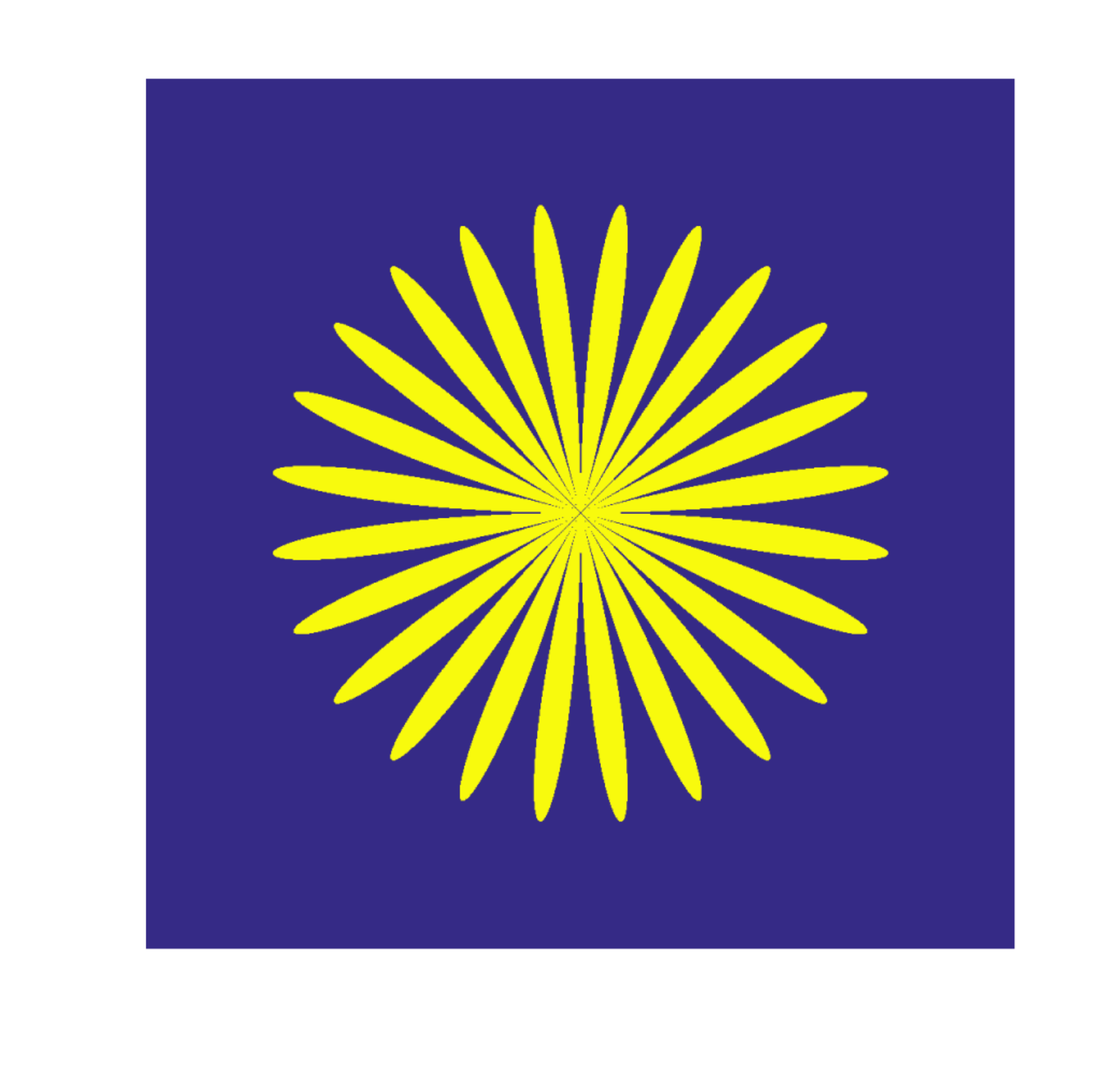}
		\caption{}\label{fig_cplx_c}
	\end{subfigure}	
\quad
		\begin{subfigure}[!t]{0.5\textwidth}
		\centering
		\includegraphics[width=0.85\linewidth]{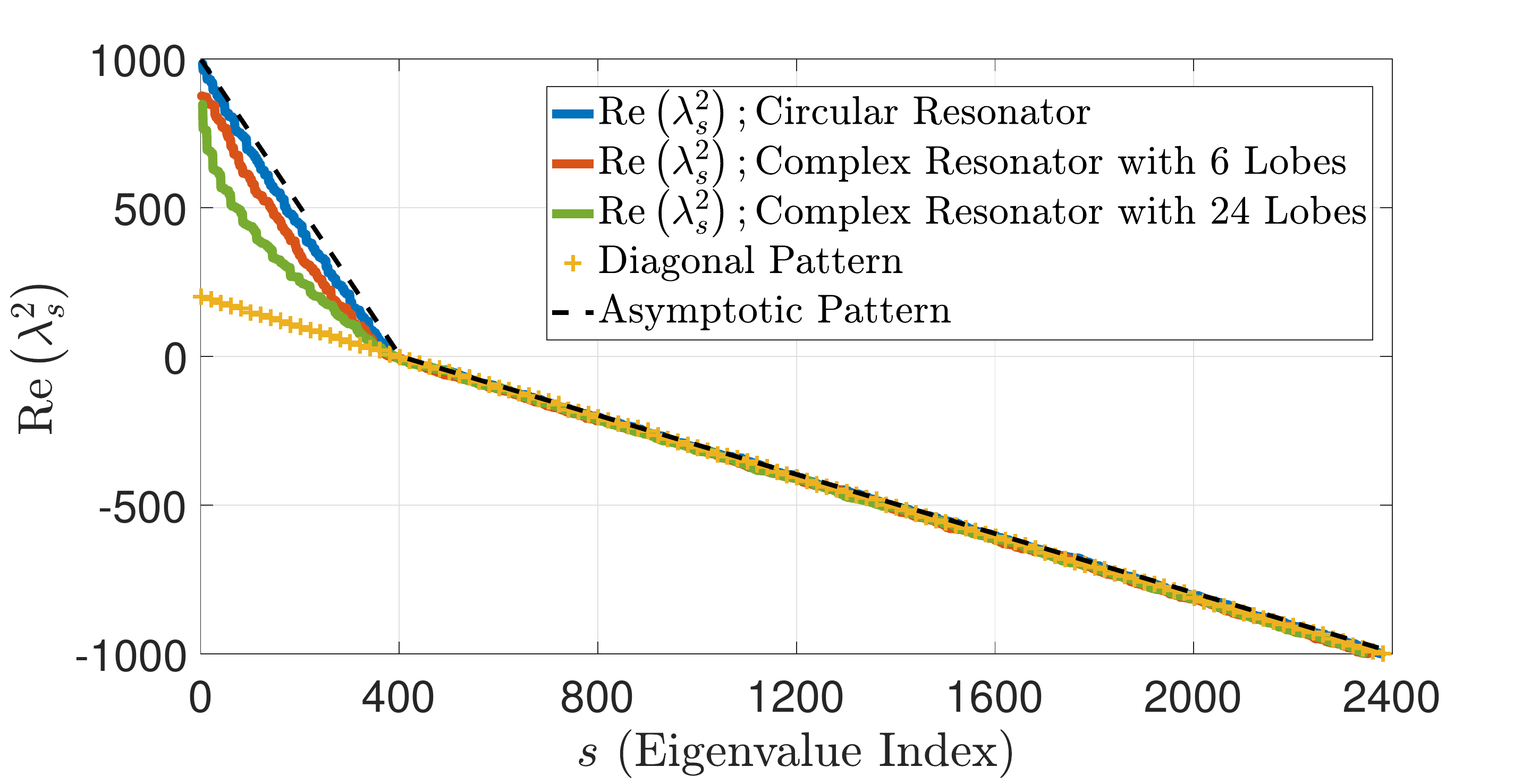}
		\caption{}\label{fig_cplx_d}		
	\end{subfigure}
\quad
		\begin{subfigure}[!t]{0.5\textwidth}
		\centering
		\includegraphics[width=0.85\linewidth]{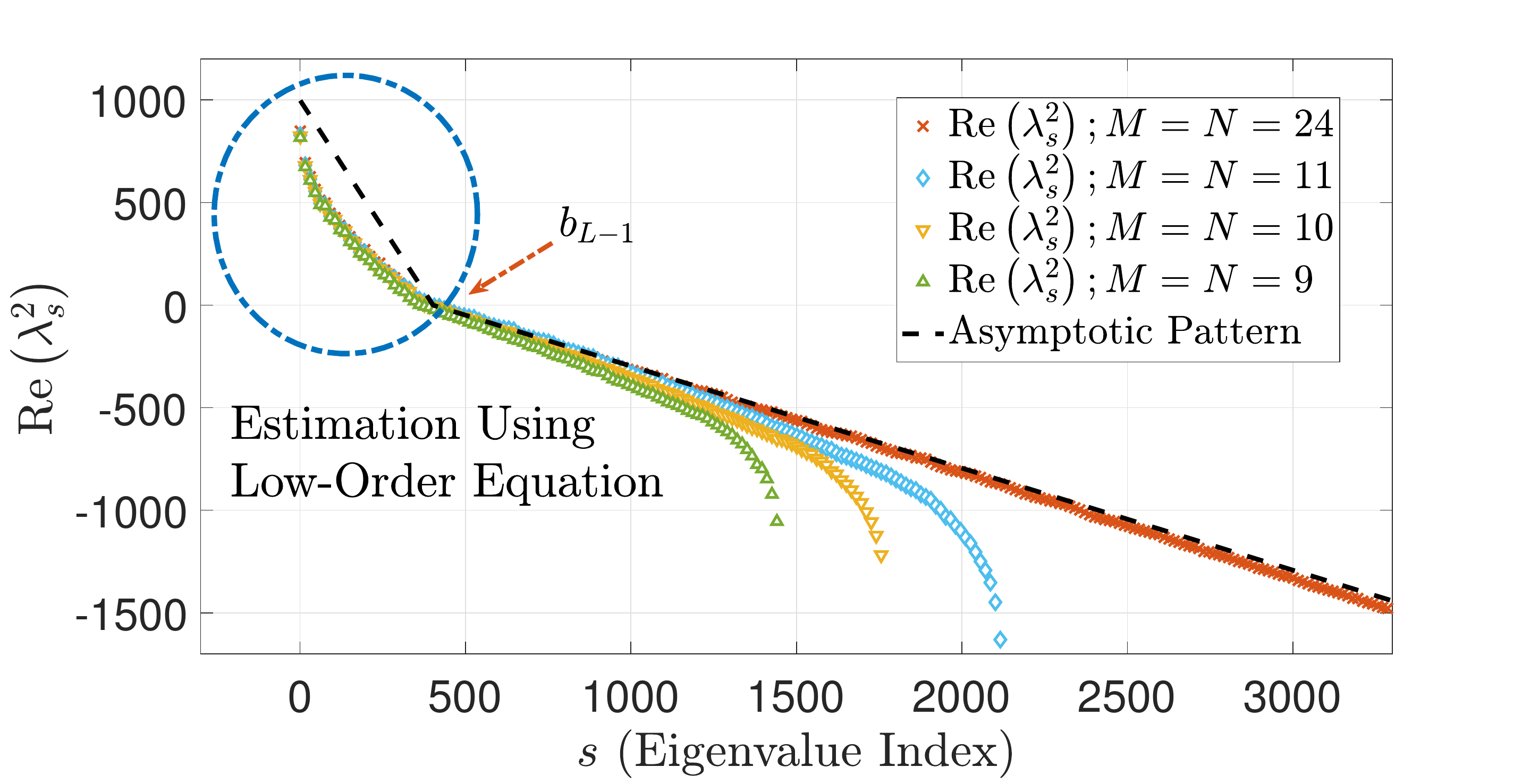}
		\caption{}\label{fig_cplx_f}		
	\end{subfigure}
\quad
		\begin{subfigure}[!t]{0.5\textwidth}
		\centering
		\includegraphics[width=0.85\linewidth]{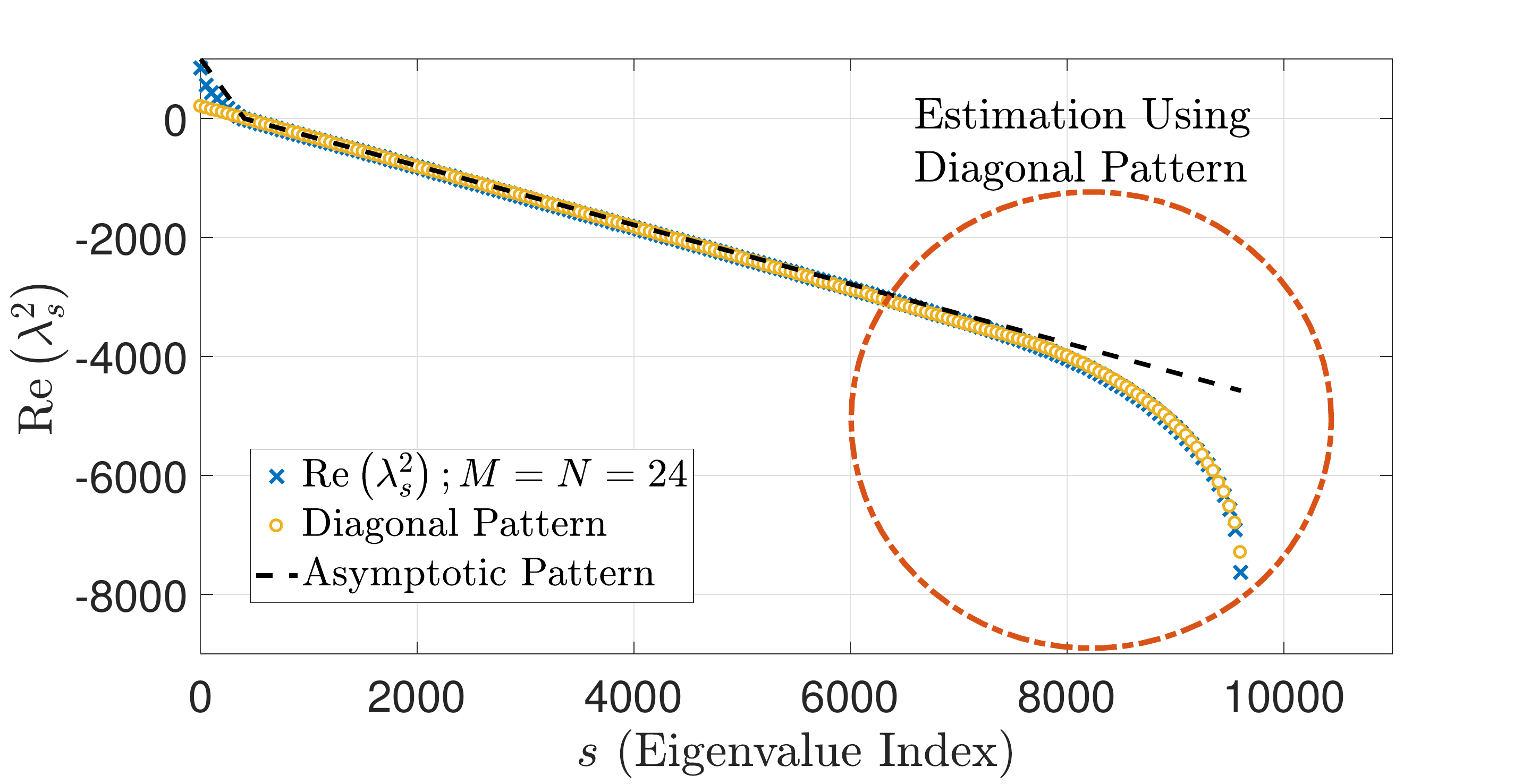}
		\caption{}\label{fig_cplx_e}		
	\end{subfigure}

	\caption{(\subref{fig_cplx_a}), (\subref{fig_cplx_b}), (\subref{fig_cplx_c}) Different resonators with similar properties $\Scale[.9]{({{\epsilon _1},{\epsilon _2}}) = ({1000,1})}$, $\Scale[.9]{{T_x} = {T_y} = 0.4}$, and $\Scale[.9]{f_1=0.2}$; (\subref{fig_cplx_d}) effect of complexity on the eigenvalue pattern; (\subref{fig_cplx_e}) estimation of complexity induced deviated eigenvalues using lower order solutions; (\subref{fig_cplx_f}) estimation of truncation induced deviated eigenvalues using the diagonal pattern.}
	\label{fig_cplx}
\end{figure}

\subsection{Complex Structures}
\label{Section: Cplx}

In this part, we show how an effective and accurate estimation of the propagation constants can be made possible, for high-contrast or complex grating structures. It was observed in Fig.~\ref{fig_par_d}, that the shape of simple resonators has little impact on the asymptotic behavior of the eigenvalues. However, as the resonator shape becomes more complex, a slight deviation at the beginning of the eigenvalue pattern emerges, i.e. for $\Scale[.85]{s<b_{L-1}}$, which can be intensified by large dielectric contrasts. This deviation, unlike the one at the tail of the eigenvalue pattern, is an estimation error and cannot be recovered directly. Nonetheless, the rest of the pattern still tends to remain indifferent to the choice of the resonator shape. This essential feature makes it possible to still find the deviated propagation constants with respect to Eq.~\ref{Eq-22}; i.e. using lower order solutions to find the first $\Scale[.9]{b_{L-1}}$ propagation constants, and the diagonal pattern to estimate the rest. To show this, three high-contrast gratings with similar $\Scale[.9]{( {{\epsilon _1},{\epsilon _2}}) = ( {1000,1})}$, $\Scale[.9]{{T_x} = {T_y} = 0.4}$, and $\Scale[.9]{f_1=0.2}$ are considered. The resonator shapes are supposed to indicate different levels of complexity: a simple circular resonator, a more complex shape with 6 lobes, and a very complex one with 24 lobes, as depicted in Fig.~\ref{fig_cplx_a}, Fig.~\ref{fig_cplx_b}, and Fig.~\ref{fig_cplx_c}, respectively. The exceedingly high contrast is intentionally chosen to reveal the utmost possible deviation. The corresponding eigenvalue patterns are plotted and compared with the asymptotic pattern in Fig.~\ref{fig_cplx_d}, indicating a deviation before the break point and a good match afterwards. It is interesting to observe that for simple resonator shapes, such as the one in Fig.~\ref{fig_cplx_a}, the asymptotic pattern still provides a relatively good estimation of the propagation constants, even with a extraordinarily high dielectric contrast. Since the deviated region remains restricted, as Eq.~\ref{Eq-21} suggests it would suffice to estimate deviated eigenvalues from a lower order solution. We have illustrated this idea in Fig.~\ref{fig_cplx_f}. For the grating of Fig.~\ref{fig_cplx_c}, the eigenvalue and the asymptotic patterns are plotted with $\Scale[.9]{M=N=24}$, resulting in a deviation. Also, the corresponding patterns with smaller orders $\Scale[.9]{M=N=11}$, $\Scale[.9]{M=N=10}$, and $\Scale[.9]{M=N=9}$ are plotted. It can immediately be recognized that for a grating as complex as that of Fig.~\ref{fig_cplx_c}, the deviated portion of eigenvalues can be estimated only with $\Scale[.9]{M=N=9}$, which will be more than a hundred times faster to obtain. Besides, the propagation constants with $\Scale[.9]{s\geqslant b_{L-1}}$ can still be estimated using the diagonal pattern, as depicted in Fig.~\ref{fig_cplx_e}, or with alternative methods.

\subsection{Regular Gratings}
\label{Section: Regul}

\begin{figure}[!ht]	
	\centering

	\begin{subfigure}{0.5\textwidth}
		\centering
		\includegraphics[width=0.87\linewidth]{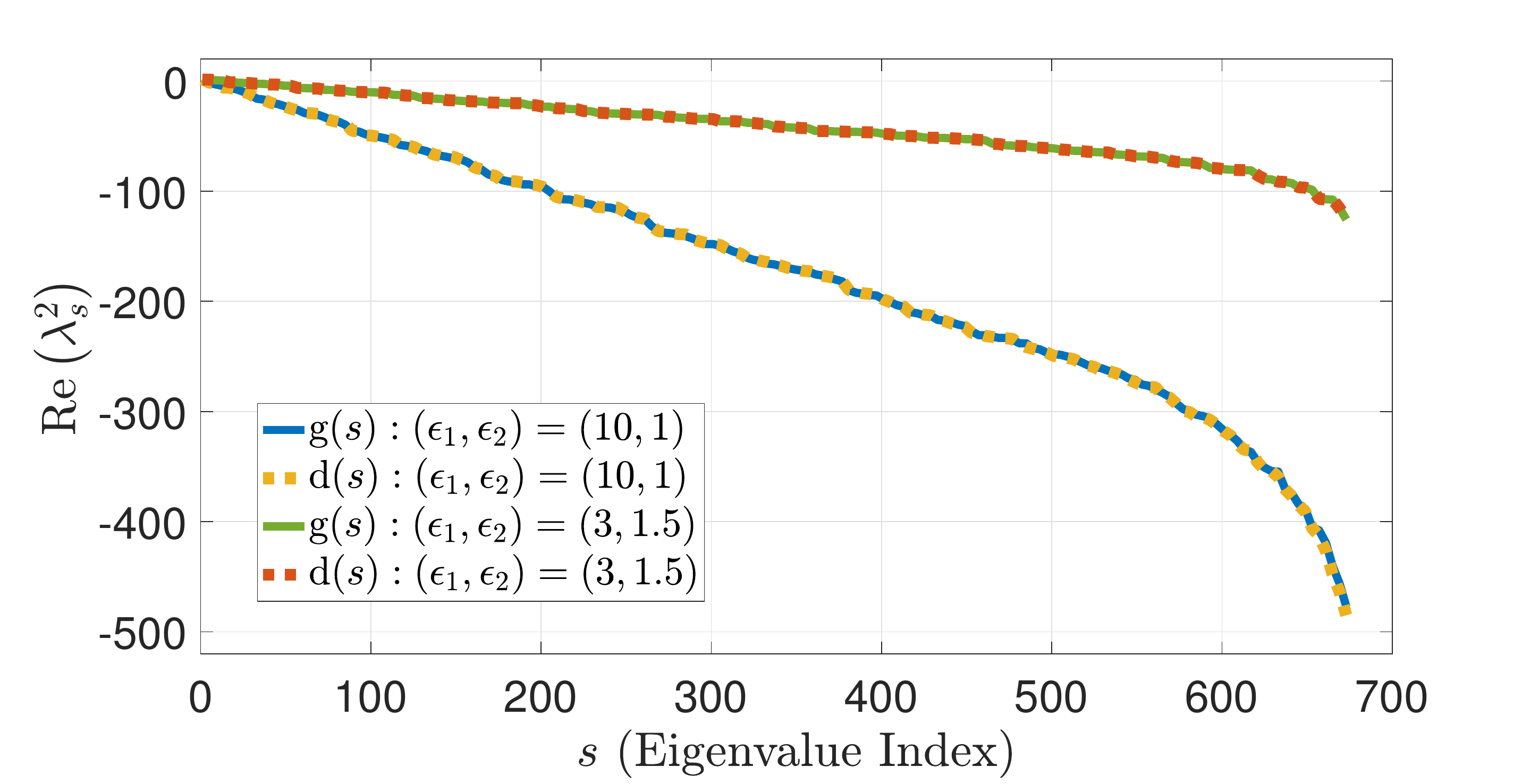}
		\caption{}\label{fig_reg_a}
	\end{subfigure}
~
	\begin{subfigure}{0.5\textwidth}
		\centering
		\includegraphics[width=.87\linewidth]{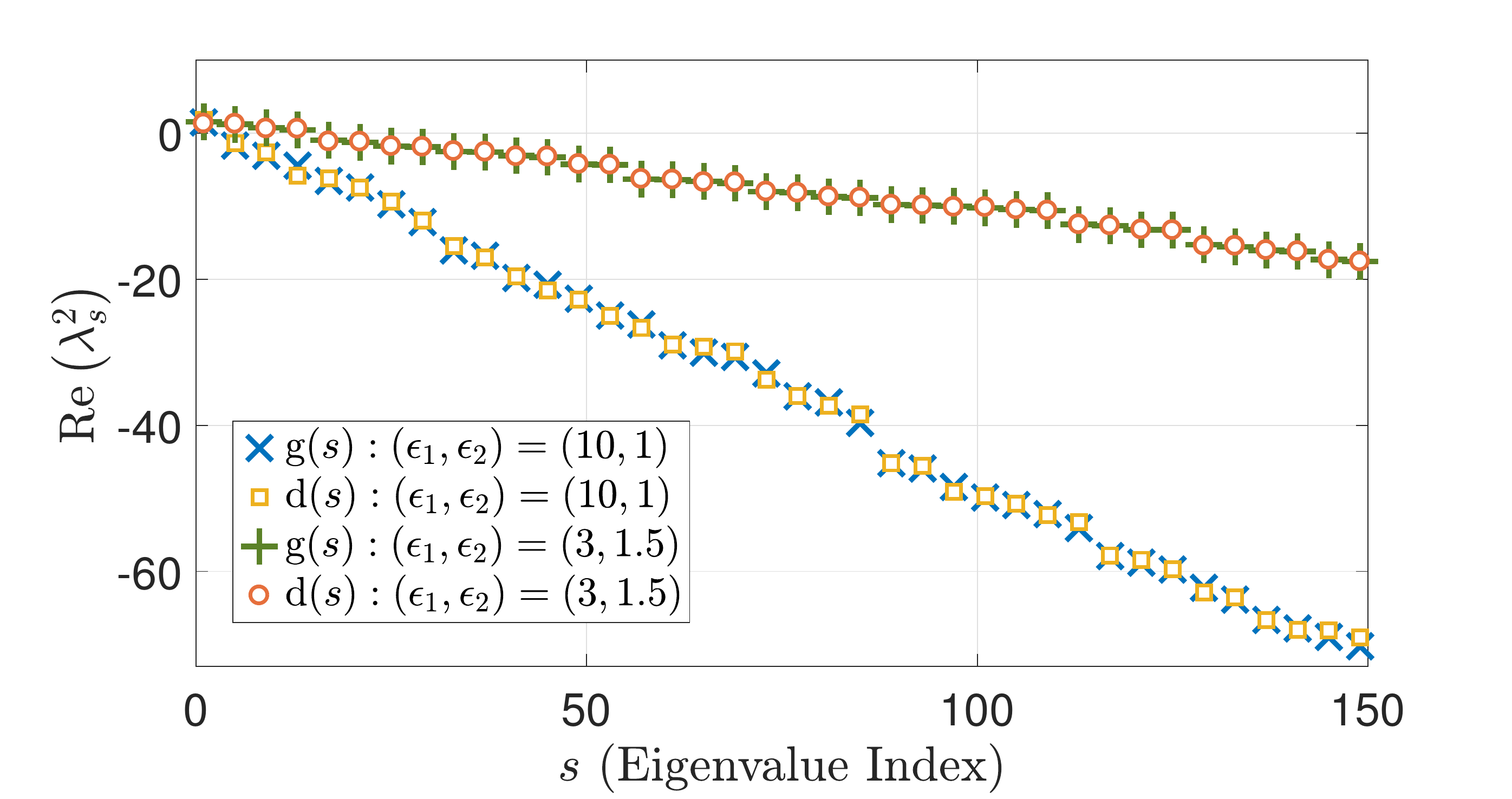}
		\caption{}\label{fig_reg_b}
	\end{subfigure}

	\caption{(\subref{fig_reg_a}) Comparison of the eigenvalue and the diagonal patterns with $\Scale[.9]{M=N=6}$, for two gratings of circular resonators, with $\Scale[.9]{({{\epsilon _1},{\epsilon _2}}) = ({10,1})}$, $\Scale[.9]{{T_x} = {T_y} = 0.4}$, $\Scale[.9]{f_1=0.15}$ for the first grating, and $\Scale[.9]{({{\epsilon _1},{\epsilon _2}}) = ({3,1.5})}$, $\Scale[.9]{{T_x} = {T_y} = 0.8}$, $\Scale[.9]{f_1=0.2}$ for the second grating; and (\subref{fig_reg_b}) a closer view of the initial points of the graphs.}
	\label{fig_regular}
\end{figure}

In the previous subsection, it was indicated how the propagation constants can be estimated even for uncannily high-contrast complex-shaped grating structures. It was all based on Eq.~\ref{Eq-22} which proposed a very general estimation scheme. Although estimation of the eigenvalues can potentially be most useful when large truncation orders are required, which is not typically the case for small dielectric constants, it is still worth providing some examples for the latter case. Particularly, since $\Scale[.85]{\lambda_s^2 = {\rm {\tilde g}}(s)+ O(1)}$ and that here $\Scale[.85]{\epsilon_k = O(1)}$, the relative error can become significant for the first few propagation constants, which happen to be the most important ones. As a result, use of $\Scale[.9]{{\rm {\tilde g}}(s)}$ might not be as useful here. However, as explained in Sub.~\ref{Subsec: Diag}, the diagonal pattern can be even more effective here. In other words, for a large category of regular gratings, Eq.~\ref{Eq-22} can be simplified to Eq.~\ref{Eq-23}. Providing two examples with ordinary numbers, we illustrate the performance of this estimation scheme. Consider two gratings of circular resonators as in Fig.~\ref{fig_cplx_a}, with $\Scale[.9]{({{\epsilon _1},{\epsilon _2}}) = ({10,1})}$, $\Scale[.9]{{T_x} = {T_y} = 0.4}$, $\Scale[.9]{f_1=0.15}$, and $\Scale[.9]{({{\epsilon _1},{\epsilon _2}}) = ({3,1.5})}$, $\Scale[.9]{{T_x} = {T_y} = 0.8}$, $\Scale[.9]{f_1=0.2}$ respectively. One can easily obtain $\Scale[.9]{b_{L-1} \approx 2.7}$ for the first grating and  $\Scale[.9]{b_{L-1} \approx 2.4}$ for the second. In fig.~\ref{fig_reg_a}, $\Scale[.9]{{\rm {g}}(s)}$ and $\Scale[.9]{{\rm {d}}(s)}$ are plotted and compared for the two gratings, and in fig.~\ref{fig_reg_b} a closer view of the patterns at the initial points is provided, both highly in consonance with the prediction of Eq.~\ref{Eq-23}.

\section{Conclusion}
\label{Section: Conc}

Modal analyses of grating structures typically give rise to infinite-sized intricate eigenvalue problems. Even after truncation, solving enormous eigenvalue equations can still be challenging due to limitations in time and memory. Adopting an innovative approach, we have proposed approximate relationships and estimation schemes for the propagations constants of a crossed grating. The propagation constants are shown to make a pattern, which can be described either in terms of the structural parameters, or as a weighted summation over the characteristic patterns of the constituent dielectrics. Moreover, based on an asymptotic result about the diagonal entries of the modal matrix, a more general estimation scheme is provided. Numerical illustration of the proposed methods is presented in the end.


\section*{Appendix}\label{Apx_A}

\renewcommand{\theequation}{A-\arabic{equation}}
\setcounter{equation}{0}  

Starting from Eq.~\ref{Eq-9}, $\Scale[.95]{\bm{P}}$ is composed of two summands and its diagonal entries are the sum of diagonal entries of the two additive terms. First, we will show that the diagonal entries of the second summand are zero in the asymptotic form. Let us designate the $p$-th diagonal entry of $\Scale[0.9]{{\bm {N_x}}^2 - \bm{N_x}{\bm{F}^{ - 1}}\bm{N_x}\bm{F} }$ and $\Scale[0.9]{{\bm {N_y}}^2 - \bm{N_y}{\bm{F} ^{ - 1}}\bm{N_y}\bm{F}}$, by $\Scale[0.95]{ d_{x,p}}$ and $\Scale[0.95]{ d_{y,p}}$, respectively. By definition, $\Scale[0.95]{ d_{x,p}}$ may be expanded as following:

\begin{equation}\label{Eq-A1}
\Scale[.85]{
\begin{array}{l}
{d_{x,p}} = [ {\bm{N_x}} ]_{p,p}^2 - {[ {\bm{N_x}}]_{p,p}}{\displaystyle\sum\nolimits_{p' = 1}^S {{{[ {\bm{N_x}}]}_{p'}}{{[ {{\bm{F} ^{ - 1}}} ]}_{p,p'}}{{[ \bm{F} ]}_{p',p}}}} = \\
\hspace{28pt} {[ {\bm{N_x}}]_{p,p}}\left( {{{[\bm{N_x}]}_p} - {\displaystyle\sum\nolimits_{p' = 1}^S {{{[ {\bm{N_x}}]}_{p',p'}}{{[ {{\bm{F}^{ - 1}}} ]}_{p,p'}}{{[\bm{F}]}_{p',p}}} }} \right).
\end{array}
}
\end{equation}

We'd like to remind here that the correspondence between the 2D periodic function $\Scale[0.9]{ \epsilon({x,y})}$ and the Fourier matrix $\Scale[.95]{\bm \epsilon}$ can be generalized to any bounded 2D periodic function defined on the unit cell. Hence, since the relationship $\Scale[0.85]{ {{\epsilon({x,y})}\times{{\epsilon({x,y})}^{-1}}= 1}}$ holds, for large enough truncated Fourier matrices we should have:

\begin{equation}\label{Eq-A2}
\Scale[.9]{
{\bm{F}}\times {\left. \bm{F}\right|_{\epsilon^{ - 1}}} \approx \bm{I}{\hspace{10pt}}\Rightarrow
{\hspace{10pt}}{{\bm{F}} ^{ - 1}} = {[ \epsilon ]^{ - 1}} \approx [ \nicefrac{1}{\epsilon } ] = {\left. \bm{F}\right|_{\epsilon^{ - 1}}}.
}
\end{equation}

This equation means the inverse of the Fourier coefficient matrix of $\Scale[0.9]{\epsilon ({x,y})}$, i.e., $\Scale[0.9]{{\bm {F}}^{-1}}$, asymptotically equals to the Fourier coefficient matrix of $\Scale[0.85]{ {\epsilon ( {x,y} )}^{-1} }$. Considering $\Scale[0.9]{ p = \nu({m,n})}$ and $\Scale[0.9]{ p' = \nu({m',n'})}$, the following equation is obtained from Eq.~\ref{Eq-A1} and Eq.~\ref{Eq-A2}, in which $\Scale[0.9]{{\epsilon ^{m,n}}}$ and $\Scale[1]{{( \nicefrac{1}{\epsilon })^{m,n}}}$ are 2D Fourier coefficients of $\Scale[.9]{\epsilon({x,y})}$ and $\Scale[0.85]{{\epsilon({x,y})}^{-1}}$, respectively:

\begin{equation}\label{Eq-A4}
\Scale[.9]{
\begin{array}{l}
{T_x}( {{N_{x0}} + {n}/{T_x}})^{-1} {d_{x,p}}= \\
\hspace{20pt}{\displaystyle\sum\limits_{m',n'} {( {m - m'} ){{({\nicefrac{1}{\epsilon } })}^{m - m',n - n'}}{\epsilon ^{m' - m,n' - n}}} }.
\end{array}
}
\end{equation}

We will prove the summation appearing in Eq.~\ref{Eq-A4} is asymptotically zero. In order to prove that, consider the Fourier series representation of $\Scale[0.85]{{\partial _x}\epsilon ( {x,y} )/{\epsilon ( {x,y} )}}$. Due to the asymptotic nature of the problem, a sufficiently smooth approximation of $\Scale[0.9]{ \epsilon ({x,y})}$ could be utilized wherever needed. Now, the expression can be expanded as following, for $\Scale[0.9]{ c_0=-j2\pi/\Lambda_x}$:

\begin{equation}\label{Eq-A5}
\Scale[.9]{
\begin{array}{l}
{\frac{{\partial _x}\epsilon({x,y})}{\epsilon( {x,y} )}} ={c_0}\left( {\displaystyle\sum\nolimits_{k,l} { {k{\epsilon ^{k,l}}{e^{ - {jkx(2\pi /{\Lambda _x})}}}{e^{ - {jly(2\pi/{\Lambda _y})} }}} } }\right) \times\\
\hspace{33pt}\left( {\displaystyle\sum\nolimits_{k',l'} { {{{( {\nicefrac{1}{\epsilon}})}^{k',l'}}{e^{ - {jk'x(2\pi /{\Lambda _x})}}}{e^{ - {jl'y(2\pi/{\Lambda _y})} }}} } }\right).
\end{array}
}
\end{equation}

Let us calculate the DC term (the Fourier coefficient with $\Scale[0.9]{m=n=0}$) of the left hand side and the right hand side of Eq.~\ref{Eq-A5} as following:

\begin{equation}\label{Eq-A6}
\Scale[.9]{
\begin{array}{l}
\Scale[.85]{\rm{LHS:}}\hspace{2pt}{\displaystyle \int\nolimits_0^{{\Lambda _y}}} {\displaystyle \int\nolimits_0^{{\Lambda _x}}{{\frac{{\partial _x}\epsilon }{{\epsilon }}}} dxdy} = {\displaystyle\int\nolimits_0^{{\Lambda _y}} {\ln \left| {\frac{{\epsilon ( {{\Lambda _x},y} )}}{{\epsilon ( {0,y})}}} \right|}dy} = 0,\\
\Scale[.85]{\rm{RHS:}}\hspace{2pt} - {c_0}{\displaystyle\sum\nolimits_k {\sum\nolimits_l {k{{( \nicefrac{1}{\epsilon } )}^{k,l}}{\epsilon ^{ - k, - l}}} }}.
\end{array}
}
\end{equation}

In the above equation, we have made use of the fact that $\Scale[0.9]{ {\epsilon({x,y})}}$ is periodic in the $x$ and $y$ directions. By equating the left and right hand side expressions, the following identity immediately emerges:

\begin{equation}\label{Eq-A7}
\Scale[.9]{
{\displaystyle\sum\nolimits_{k,l} { {k{{( \nicefrac{1}{\epsilon })}^{k,l}}{\epsilon ^{ - k, - l}}} }} =0.
}
\end{equation}

Now, by replacing $\Scale[.9]{k=m-m'}$ and $\Scale[.9]{l=n-n'}$ in Eq.~\ref{Eq-A7}, it follows from Eq.~\ref{Eq-A4} that $\Scale[0.95]{d_{x,p}=0}$. A similar procedure would result in $\Scale[.95]{d_{y,p}=0}$. Considering Eq.~\ref{Eq-9}, the above result is equivalent to $\Scale[0.85]{ \bm{\mathcal{D}}_{\epsilon}={\mathop{\delta}\nolimits} {(\bm P )^2} = {\mathop{\delta}\nolimits} {( \bm Q )^4}}$. Hence, it suffices to find the diagonal entries of $\Scale[.95]{\bm Q}$ and replicate them three more times, in order to obtain the diagonal pattern, as follows:

\begin{equation}\label{Eq-A8}
\Scale[.9]{
\begin{array}{l}
\delta\left( \bm{Q} \right) = {\{ {{\epsilon ^{0,0}} - {{\left( {{N_{x0}} + {m}/{T_x}} \right)}^2} + {{\left( {{N_{y0}} + {n}/{T_y}} \right)}^2}} \}_{m,n}},\\
{\epsilon ^{0,0}} ={\displaystyle\int_0^{{\Lambda _y}}} {\displaystyle\int_0^{{\Lambda _x}} {\epsilon \left( {x,y} \right)dxdy} } ={\epsilon _{avg}} = {\displaystyle\sum\nolimits_{i = 1}^L {{f_i}{\epsilon _i}}}.
\end{array}
}
\end{equation}

It follows that the diagonal entries are the same as the eigenvalues of a uniform layer with $\Scale[0.9]{ {\epsilon _1} = {\epsilon _{avg}}}$.


%

\end{document}